\documentclass[acmsmall, nonacm, screen]{acmart}

\usepackage{comment}
\usepackage{listings}
\usepackage{placeins}
\usepackage{array}
\usepackage[utf8]{inputenc} 
\usepackage[english]{babel}
\usepackage{color, colortbl}
\usepackage{csquotes}
\usepackage{enumitem}

\usepackage[shortcuts]{extdash} 
\usepackage[T1]{fontenc} 
\usepackage{booktabs}
\usepackage{threeparttable} 
\usepackage{graphicx}
\usepackage{listings}
\usepackage{hyphenat}
\usepackage{multirow}
\usepackage{seqsplit}
\usepackage{siunitx}
\usepackage{subcaption}
\usepackage{tabularx}
\usepackage{xcolor}
\usepackage{xspace}
\usepackage{url}
\usepackage{balance}
\usepackage{pgfplots}
\usepackage{enumitem}
\usepackage[T1]{fontenc}
\usepackage{listings}
\usepackage{amsmath}
\usepackage{caption}
\usepackage{subcaption}
\usepackage{xcolor}
\usepackage{tikz}

\usetikzlibrary{shapes, arrows.meta, positioning}
\definecolor{codegreen}{rgb}{0,0.6,0}
\definecolor{codegray}{rgb}{0.5,0.5,0.5}
\definecolor{codeblack}{rgb}{0,0,0}
\definecolor{codepurple}{rgb}{0.58,0,0.82}

\definecolor{custompurple}{RGB}{115, 3, 252}
\definecolor{customyellow}{RGB}{252, 227, 3}
\definecolor{backcolour}{rgb}{0.95,0.95,0.92}
\definecolor{lightgray}{rgb}{0.9,0.9,0.9}

\lstdefinestyle{mystyle2}{  
    commentstyle=\color{codegreen},
    keywordstyle=\color{magenta},
    numberstyle=\tiny\color{codegray},
    stringstyle=\color{codepurple},
    basicstyle=\fontfamily{put}\selectfont\tiny,
    breakatwhitespace=false,         
    breaklines=true,                 
    captionpos=b,                    
    keepspaces=true,           
    numbers=left,
    escapeinside={(*@}{@*)},
    numbersep=5pt,          
    showspaces=false,                
    showstringspaces=false,
    showtabs=false,                  
    tabsize=2,
    xrightmargin=3pt,
    escapechar=@,
}

\lstdefinestyle{bashstyle}{
    language=bash,
    basicstyle=\fontfamily{put}\selectfont\tiny,
    keywordstyle=\color{blue},
    commentstyle=\color{gray},
    stringstyle=\color{orange},
    showstringspaces=false,
    numbers=left,
    numberstyle=\tiny\color{codegray},
    numbersep=5pt,
    breaklines=true,
    xleftmargin=5pt,
    xrightmargin=3pt,
    breakatwhitespace=true,
    frame=single,
    rulecolor=\color{black},
    backgroundcolor=\color{white},
    extendedchars=true,
    tabsize=4,
    escapechar=@,
}
\addto\extrasenglish{%

}
\newcolumntype{C}[1]{>{\centering\arraybackslash}p{#1}}
\newcommand{\useconcmath}{\fontfamily{ccr}\selectfont}

\newcommand{\sysname}{\textsc{WasmChecker}\xspace}

\title{Reusing Legacy Code in WebAssembly: Key Challenges of Cross-Compilation and Code Semantics Preservation}

\makeatother

\author{Sara Baradaran}
\affiliation{%
  \institution{University of Southern California}
  \country{USA}}
\email{sbaradar@usc.edu}

\author{Liyan Huang}
\affiliation{%
  \institution{University of Southern California}
  \country{USA}}
\email{liyanhua@usc.edu}

\author{Mukund Raghothaman}
\affiliation{%
  \institution{University of Southern California}
  \country{USA}}
\email{raghotha@usc.edu}

\author{Weihang Wang}
\affiliation{%
  \institution{University of Southern California}
  \country{USA}}
\email{weihangw@usc.edu}

\begin{document}

\begin{abstract}
WebAssembly (Wasm) has emerged as a powerful technology for executing high-performance code and reusing legacy code in web browsers. With its increasing adoption, ensuring the reliability of WebAssembly code becomes paramount. In this paper, we investigate how well WebAssembly compilers fulfill code reusability. Specifically, we inquire (1) what challenges arise when cross-compiling a high-level language codebase into WebAssembly and (2) how faithfully WebAssembly compilers preserve code semantics in this new binary. Through a study on 115 open-source codebases, we identify the key challenges in cross-compiling legacy C/C++ code into WebAssembly, highlighting the risks of silent miscompilation and compile-time errors. We categorize these challenges based on their root causes and propose corresponding solutions. We then introduce a differential testing approach, implemented in a framework named \sysname, to investigate the semantics equivalency of code between native x86-64 and WebAssembly binaries. Using \sysname, we provide a witness that WebAssembly compilers do not necessarily preserve code semantics when cross-compiling high-level language code into WebAssembly due to different implementations of standard libraries, unsupported system calls/APIs, WebAssembly's unique features, and compiler bugs. Furthermore, we have identified 11 new bugs in the Emscripten compiler toolchain, all confirmed by Emscripten developers. As proof of concept, we make our framework and the collected dataset of open-source codebases publicly available.
\end{abstract}

\maketitle
\section{Introduction}
\label{Introduction}
WebAssembly, also known as Wasm, is a statically typed binary instruction format that serves as a portable compilation target for high-level programming languages such as C, C++, Rust, and Go \cite{WebAssemblyPLDI}. It has been designed to provide a compact, platform-independent bytecode that can be executed within a web browser at a speed close to native binary execution. This feature encourages web developers to use WebAssembly for performance-intensive applications, such as gaming \cite{battagline2019hands}, virtual reality \cite{10.1145/3618257.3624833, khomtchouk2021webassembly}, and audio/video processing \cite{10335388, kaluva2020webassembly}.

Since WebAssembly's inception in 2017, it has gained significant adoption and support from vendors of major browsers, including Chrome, Firefox, Safari, and Edge \cite{can_i_use_wasm}. WebAssembly ecosystem is currently expanding beyond the browser into a multitude of domains, from edge computing \cite{9283720, 10034550} to the Internet of Things (IoT) \cite{fi15080275, 9156135, 10.1145/3498361.3538922} and smart contracts \cite{10.1145/3533767.3534218, 272292, WASMOD}. This expansion has led to the development of various compilers for translating high-level language programs into low-level WebAssembly code \cite{emscripten, binaryen, AssemblyScript, Cheerp, Asterius, Wasm-Bindgen}.

Although achieving higher execution speeds is a significant motivation for WebAssembly development, this is not the only objective. WebAssembly also aims to facilitate code reuse in web development by integrating vast existing codebases and libraries written in various languages into web applications instead of writing JavaScript code from scratch. This assumes that programmers are able to cross-compile high-level source code into WebAssembly to use as a part of a web application. However, recent studies indicate that when a high-level code is compiled and run as a Wasm binary, the program's behavior may not be uniformly preserved due to the immaturity of WebAssembly compilers and implementation discrepancies among different WebAssembly runtimes \cite{9724846, stievenart2022security, 255318}. Specifically, these studies highlight that unexpected vulnerabilities might become exploitable when a program is compiled into Wasm binary. Such vulnerabilities stem from differences in WebAssembly's memory model compared to native binaries and the lack of security protection mechanisms, such as stack canaries, in WebAssembly compilers.

In the same direction, study \cite{9678776} shows that WebAssembly compilers, similar to compilers of other programming languages, are not bug-free. Developers of WebAssembly compilers face unique challenges that can introduce buggy behaviors specific to this new binary. For instance, while C/C++ supports a fully synchronous execution model, browsers intrinsically execute JavaScript code asynchronously. Therefore, WebAssembly compiler developers must ensure that synchronous operations in C/C++ programs are properly ported to the asynchronous browser environment. Otherwise, asynchronous execution of such operations may result in bugs and unexpected program behaviors. 

Considering these challenges, if WebAssembly is to be a viable method for using legacy code in web applications, the developer community must first ensure that the original code's semantics remains consistent when compiled into WebAssembly. Otherwise, unexpected behaviors in WebAssembly code can introduce new bugs into web applications.

\textit{\textbf{This work.}} In this paper, we aim to understand how effectively fledgling WebAssembly compilers enable code reusability. Unlike previous studies \cite{9724846, stievenart2022security, 255318}, which exclusively examine the security behavior of programs cross-compiled into WebAssembly, we target the \textit{functional behavior} of code. Our inquiry raises the following two important questions: 

\begin{itemize}[leftmargin=*, itemindent=9pt]
    \item \textbf{RQ1:} What \textit{challenges} arise when cross-compiling high-level language code into WebAssembly?
    
    \item \textbf{RQ2:} 
    Can WebAssembly compilers preserve \textit{source code semantics} in the resulting Wasm binaries? 
\end{itemize}

In the first phase of our study, we focus on RQ1 to investigate whether cross-compiling C/C++ applications into WebAssembly is straightforward and to identify the challenges that arise without changing the original codebases and build commands. We collected 115 open-source C/C++ projects and manually built them into x86-64 and WebAssembly binaries. We observed that the compilation and build process in WebAssembly is not always as straightforward as in native binary. In many cases, challenges such as unsupported APIs, incorrect compiler settings, and dependency issues cause the compilation and build process in WebAssembly to fail. We divide these failures into 6 categories based on their root causes and discuss corresponding solutions. Insights gained from this phase lay the groundwork for a systematic approach to address RQ2 in the second phase.

To answer RQ2, we use a test-driven approach to evaluate code semantics in native (e.g., x86-64) and WebAssembly binaries. Large-scale projects typically contain hundreds of rich test cases usable for regression testing, refactoring, and code maintenance. By leveraging such test cases, we propose \sysname, a framework for differential testing between C/C++ applications and their WebAssembly counterparts. Given a C/C++ program and a set of test cases evaluating its functional correctness, \sysname first builds the source code alongside the test cases into two distinct binaries, x86-64 and WebAssembly. Then, it compares the functional behavior of the program in two binaries by executing executable tests and comparing their outcomes, thereby reporting potential dissimilarities between them.

In order to determine how faithfully WebAssembly compilers preserve source code semantics in Wasm binaries, we deployed \sysname on 135 open-source C/C++ projects, comprising 34,480 test cases. This included 115 projects studied in the first phase and a random sample of 20 new projects, distinct from the initial 115 projects, to minimize evaluation bias. \sysname showed that out of 5,862 executable tests, 226 exhibited differing outcomes when executed in different binaries. We analyzed the reported discrepancies and their underlying root causes to highlight the disparities between WebAssembly compilers and traditional ones. Additionally, we identified 11 new bugs within Emscripten \cite{emscripten}, the most widely-used WebAssembly compiler, which compiles C/C++ programs into WebAssembly. We have documented and reported these new bugs, all of which have been confirmed by Emscripten developers.

\noindent
\textit{\textbf{Contributions.}} This paper makes the following contributions:

\begin{itemize}[leftmargin=*]
    \item We identify and categorize the key challenges in cross-compiling high-level language code into Wasm binaries, upon which we propose an automated approach for building C/C++ codebases in WebAssembly (Section~\ref{sec:RQ1} and Section~\ref{sec:RQ2-A}).
    \item We present \sysname, a differential testing framework for analyzing semantics equivalency between C/C++ programs compiled into native x86-64 binaries and their WebAssembly counterparts (Section~\ref{sec:RQ2-B}).
    \item We evaluate \sysname on 135 open-source projects with 34,480 test cases from GitHub, showcasing the gaps in code semantics translation between WebAssembly compilers and traditional C/C++ compilers (Section~\ref{sec:RQ2-C} and Section~\ref{sec:RQ2-D}). 
    \item We identified 11 new bugs within the Emscripten compiler, all of which have been confirmed by Emscripten developers.
    \item We make the collected dataset of open-source codebases, the scripts to reproduce the results, and the source code of \sysname publicly available to encourage future work in this area.
\end{itemize}

\section{Methodology}
\label{sec:Methodology}
\noindent
To understand how effectively WebAssembly compilers realize legacy code reuse, a two-phase study has been designed. The first phase of our study focuses on identifying and categorizing potential challenges one might encounter when cross-compiling high-level language codebases into WebAssembly using currently available compilers. To do this, we attempted to build codebases of 115 open-source projects collected in a dataset in both x86-64 binary and WebAssembly. These projects cover a wide range of real-world codebases with diverse code constructs. The methodology for data collection and selection criteria are detailed in Section~\ref{sec:dataset}. Here, we seek to discover common scenarios where, unlike compiling code into native binary, cross-compiling code into WebAssembly fails and requires changes in the codebase or build commands. We classify observed failures into 6 categories and specify if each needs a significant amount of code changes or whether it could be resolved by applying minor changes to the build commands. 

The second phase of our study focuses on understanding how well WebAssembly compilers preserve code semantics when compiling into WebAssembly. We design and implement a differential testing framework by taking advantage of open-source test cases to compare the functional behavior of a given codebase across two different binaries. Our framework first systematically builds the codebase, including the program's source code and test cases, into two different binaries. Then, it executes executable tests in these binaries and captures their outcomes to subsequently compare and report potential discrepancies. Using this approach, we provide a witness that cross-compiling C/C++ code into WebAssembly does not necessarily port the exact code semantics. 

\begin{figure}[t!]
    \centering
\includegraphics[width=\linewidth]{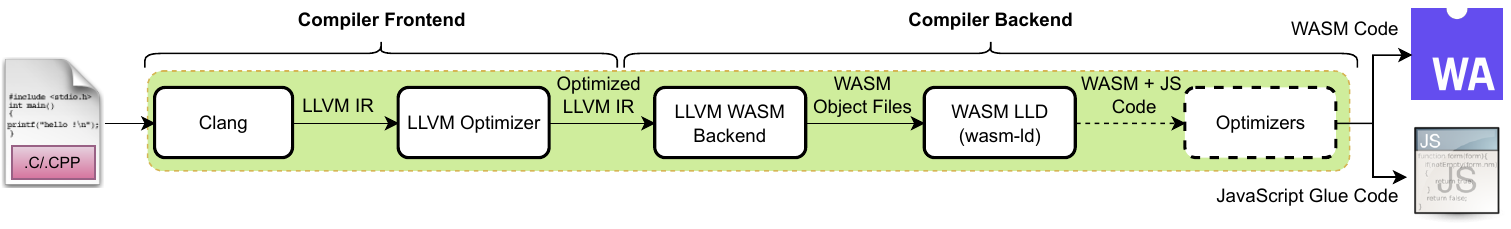}
\vspace{-2.4em}
    \caption{Emscripten compiler architecture.} 
    \label{fig:emscripten}
    \vspace{-0.7em}
\end{figure}

\textit{\textbf{Compiler toolchain.}}
In this study, we use Emscripten, the most widely-used compiler toolchain for WebAssembly \cite{9678776}, to build projects in Wasm binaries and the GNU Compiler Collection (GCC) \cite{gcc} for x86-64 binaries. Fig.~\ref{fig:emscripten} depicts the internal structure of Emscripten as an LLVM-based compiler. Emscripten first takes C/C++ code and uses Clang as the frontend to analyze the syntax and semantics of the source code. Clang translates the parsed code into LLVM Intermediate Representation (IR). The IR code is then optimized for size and efficiency. In Emscripten's backend, the LLVM WebAssembly backend translates the optimized IR into WebAssembly code, producing WebAssembly object files. Finally, the wasm-ld linker combines multiple generated object files with required libraries to create a single executable or shared library. 

Since the WebAssembly module produced by Emscripten is not standalone and cannot directly interact with Web APIs, Emscripten generates a JavaScript glue code to instantiate the WebAssembly module. This glue code is responsible for importing necessary functions, allocating memory, emulating the local file system, and translating errors into exceptions or meaningful messages.

Passing optimization flags to Emscripten at compile time adds a multi-level optimization process to the linking phase to improve the performance and reduce the executable code size (dotted-line step in Fig.~\ref{fig:emscripten}). Binaryen \cite{binaryen} is used for optimizations such as code inlining at the level of WebAssembly modules. Emscripten's JS optimizer is applied to optimize the JavaScript glue code. The compiler may also optimize the combined Wasm and JavaScript code by minifying imports and exports between the two and eliminating dead code \cite{optimization}.

\subsection{Data Collection}
\label{sec:dataset}
To gain a comprehensive insight into the challenges of cross-compiling high-level codebases into WebAssembly, we collected a dataset of 115 open-source C/C++ projects. This dataset includes open-source libraries with a wide range of functionalities, such as parsing documents in various formats (e.g., JSON, YAML, XML), calculating mathematical functions, and implementing new data structures. We specifically targeted CMake-based projects on GitHub that meet three criteria: 

\begin{enumerate}[leftmargin=*]

\item Projects having test cases generated by developers or utilizing testing frameworks (e.g., Catch2 \cite{Catch2}, GoogleTest \cite{Gtest}), evidenced by a test directory in the repository (e.g., {\useconcmath test}/{\useconcmath tests}). 

\item Projects including a CMake build script (i.e., {\useconcmath CMakeLists.txt}) in the root directory, since Emscripten provides {\useconcmath emcmake} utility that facilitates the build process in WebAssembly. 

\item Projects with over 50 stars, which generally suggest higher code quality and maturity. 
\end{enumerate}

We used the GitHub Search API \cite{githubapi} to collect repositories that meet our criteria, resulting in an initial dataset of 180 C/C++ projects. We conducted a two-phase refinement on the collected dataset. First, by reading the documentation of each project, we filtered out 46 codebases explicitly designed for platform-specific tasks, e.g., efficient memory management for Windows applications. This ensures that we focus on projects designed for general tasks that might be demanded on any platform, including the web environment. Second, we built all the projects in native x86-64 binaries using GCC and CMake and ruled out 19 projects for which the build process failed (mostly due to outdated build scripts incompatible with the current project's version).

Table~\ref{tab:dataset} summarizes the final dataset. From left to right, columns represent the functionality of codebases, the number of projects/repositories collected for each category, total lines of code, total number of test cases, and executable tests. In this study, executable tests refer to files that execute one or more test cases. When discussing executable tests in the context of native binaries, we refer specifically to ELF files that execute tests on a native platform, in our case, a server running Ubuntu 22.04 LTS. Conversely, in the context of WebAssembly, executable tests refer to JavaScript files that instantiate and manage WebAssembly modules produced by cross-compiling test cases.

\begin{table}[t]
  \caption{Overview of the collected dataset of open-source projects.}
  \label{tab:dataset}\small
  \begin{tabular}{p{7.1cm}crrr}
    \toprule
    \textbf{Functionality} & \textbf{Projects}\tnote{a} & \textbf{KLOC} & \textbf{Test Cases} & \textbf{Exe. Tests}\tnote{b}\\
    \midrule
    Utility$/$general-purpose (formatting, logging, etc) & 40 & 1,302\hspace{1mm} & 14,308\hspace{1mm} & 1,929\hspace{1mm} \\
    Data parsing$/$serialization$/$management & 25 & 1,436\hspace{1mm} & 5,529\hspace{1mm} & 712\hspace{1mm} \\
    Algorithms and data structures & 23 & 795\hspace{1mm} & 3,684\hspace{1mm} & 2,767\hspace{1mm} \\
    Mathematical$/$graphical$/$numerical computations & 20 & 4,830\hspace{1mm} & 2,339\hspace{1mm} & 550\hspace{1mm} \\
    Networking and web & 7 & 2,055\hspace{1mm} & 3,042\hspace{1mm} & 435\hspace{1mm} \\
    \hline
    \rowcolor{lightgray}Total & 115 & 10,420\hspace{1mm} & 28,902\hspace{1mm} & 6,393\hspace{1mm} \\
  \bottomrule
\end{tabular}
\end{table}

\subsection{Building the Codebases}
\label{sec:building-codebases}
Using the projects' build scripts, we attempted to build each project collected in the dataset in both x86-64 and WebAssembly binaries. We successfully built 81 (70\%) projects into WebAssembly using the same commands as for x86-64 binaries. For example, Fig.~\ref{fig:build} shows sample commands on the left building the yaml-cpp project \cite{yaml-cpp}, a YAML parsing library, into x86-64 binary, while commands on the right build it into WebAssembly. However, for the remaining 34 projects, unlike the build process in x86-64, building these projects into WebAssembly introduces errors that break the build procedure. We analyzed these errors to understand the root causes of build failures in WebAssembly. Such root causes hinder the reuse of legacy C/C++ code in WebAssembly or at least make it challenging. In the next section, we categorize these failures by their root causes.

\begin{figure}[b]
    \centering
    \begin{subfigure}[b]{0.46\linewidth}
    \centering
    \begin{lstlisting}[style=bashstyle, language={},xleftmargin=4pt, numbers=none]
$ mkdir build && cd build
$ cmake @{$\tikz \fill (0,0) circle (0.5pt);$}@@{\thinspace}@@{$\tikz \fill (0,0) circle (0.5pt);$}@ @{\textendash}@DYAML_CPP_BUILD_TESTS=ON
$ cmake @{\textendash\thinspace\textendash}@build @{$\tikz \fill (0,0) circle (0.5pt);$}@
\end{lstlisting}
    \end{subfigure}
    \hfill
    \begin{subfigure}[b]{0.53\linewidth}
    \begin{lstlisting}[style=bashstyle, language={},xleftmargin=0pt, numbers=none]
$ mkdir build && cd build
$ emcmake cmake @{$\tikz \fill (0,0) circle (0.5pt);$}@@{\thinspace}@@{$\tikz \fill (0,0) circle (0.5pt);$}@ @{\textendash}@DYAML_CPP_BUILD_TESTS=ON
$ emmake cmake @{\textendash\thinspace\textendash}@build @{$\tikz \fill (0,0) circle (0.5pt);$}@
\end{lstlisting}
    \end{subfigure}
    \vspace{-0.5em}
    \caption{Commands for building a sample codebase in x86-64 and their equivalent for building in WebAssembly.}
    \vspace{-0.5em}
    \label{fig:build}
\end{figure}

\section{RQ1: Challenges in Migrating Codebases to WebAssembly}
\label{sec:RQ1}
To answer RQ1 and explore how straightforward it is to cross-compile C/C++ code into WebAssembly, we investigate the main challenges encountered when building real-world projects into WebAssembly using Emscripten. These challenges lead to either compilation or linking errors, which can be divided into 6 categories based on their underlying reasons. Table~\ref{tab:build-results} shows the distribution of build challenges in our dataset. Note that one may need to address multiple challenges when cross-compiling a C/C++ codebase into WebAssembly. 

\textit{\textbf{Undefined symbols.}} If the code refers to functions or variables not defined within the program or linked libraries, Emscripten will generate an \textit{undefined symbol} error. This error indicates the object file or library containing the symbol definition is not properly linked. An example frequently observed is an error with the message {\useconcmath undefined symbol:\_\_stack\_chk\_guard}, resulting from the lack of support for Stack Smashing Protection (SSP), a security mechanism to detect buffer overflow \cite{cowan1998stackguard}. Upon calling a function, the SSP inserts a value, called stack canary, into the call stack before inserting local variables of the callee function. This value is stored in variable {\useconcmath \_\_stack\_chk\_guard}. The SSP also adds instructions before the function’s return, checking whether the canary value has remained intact. If the canary value fails to pass the sanitary check, a stack check failure function, namely {\useconcmath \_\_stack\_chk\_fail}, is called, which terminates the program to prevent stack-based attacks.

\begin{figure}[t]
    \centering    \includegraphics[width=0.75\linewidth]{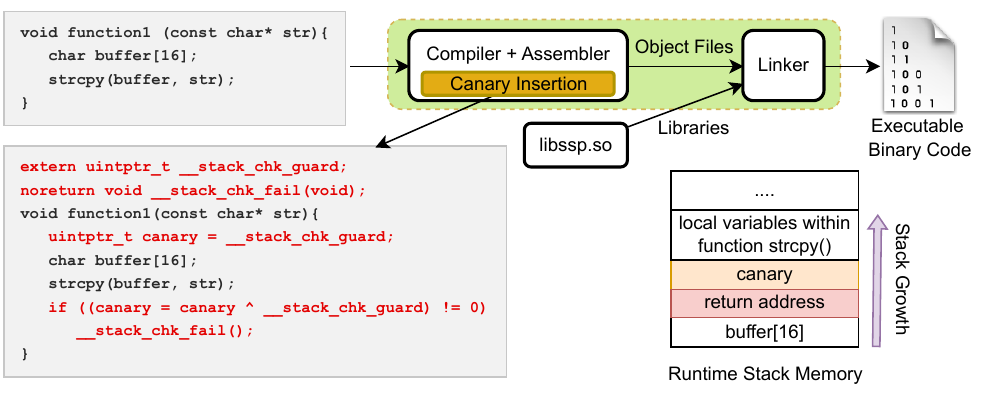}
    \vspace{-1.0em}
    \caption{Stack smashing protection in C/C++ compilers.}
    \label{fig:canary}    
    \vspace{-1em}
\end{figure}

Traditional C/C++ compilers like GCC and Clang have an option for enabling SSP (e.g., {\useconcmath -fstack-protector}), which adds the necessary stack protector support when compiling C/C++ code. As depicted in Fig.~\ref{fig:canary}, this option declares the external variable {\useconcmath \_\_stack\_chk\_guard} and the function {\useconcmath \_\_stack\_chk\_fail} in the code alongside defining a variable {\useconcmath canary}. It also adds a code section to check the canary value and call the failure handler function. The compiler then produces object files for the resulting code, which are later combined by the linker and linked against the stack protector runtime library (e.g., {\useconcmath libssp} in GCC) in order to resolve references to external symbols.

Since Emscripten internally takes Clang to compile C/C++ code into LLVM IR, it supports most Clang compiler options, including those that enable SSP. However, Emscripten does not offer a runtime library to support stack protection. As a result, the wasm-ld linker introduces an error when a program tries to link an object file with stack protection calls inserted. 

In addition to the lack of support for the SSP, there are other limitations in Emscripten, which result in undefined symbol errors. For example, since WebAssembly is executed in a Virtual Machine (VM), the functionality of POSIX APIs, dependent on the operating system or hardware access, must be emulated to perform without real access to system resources. Currently, Emscripten does not provide full support for POSIX APIs, albeit it partially emulates certain APIs, such as those for working with file systems. Hence, an undefined symbol error may arise when compiling code containing calls to unsupported API functions.

Although WebAssembly's official documentation \cite{portability} leaves it to the compiler to adapt the standard interfaces to the host environment’s available imports, the current versions of WebAssembly compilers are not mature enough to thoroughly perform this adaption. In the case of using Emscripten, unsupported API calls must be omitted from the code to make it portable to WebAssembly.

\textit{\textbf{Missing third-party libraries.}} 
Reusing third-party libraries as dependencies is a common practice to save time and manpower in software development \cite{10.1145/3551349.3560432}. Large-scale codebases sometimes use third-party libraries such as Boost \cite{boost}, a collection of libraries extending the C++ standard library with functions and structures for linear algebra, random number generation, etc. When building such codebases in WebAssembly, CMake is unable to locate the header files for third-party libraries, thereby raising errors. To build codebases that use third-party libraries, one needs to first build these libraries in WebAssembly and link them with the main program. Only a limited number of useful libraries have already been ported to WebAssembly, which reside in emscripten-ports repositories \cite{emports}. Each ported library can be used via a specific compiler flag. For example, to compile a codebase that uses the Boost library, the compiler option {\useconcmath\footnotesize USE\_BOOST\_HEADERS} should be set. This instructs Emscripten to fetch the Boost library from the remote server, set it up, build it locally, and then link it against the project.

\textit{\textbf{Target-dependent warnings + Werror.}} The compiler option {\useconcmath-\small{W}error} is used to treat all warnings as errors. When this flag is enabled, any warning generated during the compilation process causes the compiler to halt and prevent it from generating executable code. In our dataset, some projects enforce a stricter code standard by enabling {\useconcmath-\small{W}error} in their build scripts. This can result in build failures in WebAssembly, even though the codebase can still be compiled into a native binary. Due to different compilation targets (i.e., x86-64 and Wasm), a compiler may generate warnings while the other compiles the same code without introducing any warnings. We call such warnings as \textit{target-dependent}. For example, the flag {\useconcmath-mtune} in traditional compilers instructs the compiler to optimize the generated code's performance for a particular processor. Setting {\useconcmath-mtune=native} optimizes the binary code for the processor on which the compilation takes place without restricting it to run only on that processor. Since Emscripten cross-compiles code into platform-independent WebAssembly, in case of using {\useconcmath-mtune=native} as a compiler option, Emscripten ignores it, and an \textit{unused compilation argument} warning is produced. This warning, when paired with the enabled {\useconcmath-\small{W}error} option, results in a build failure. In such cases, removing {\useconcmath-\small{W}error} or {\useconcmath-mtune} allows the compilation to proceed without affecting the generated code's functionality. 

\textit{\textbf{Architecture- and platform-specific code.}} Codebases that use architecture- or platform-specific code fail to be built in WebAssembly using Emscripten. For example, C/C++ code with architecture-specific inline assembly (e.g., an {\useconcmath asm}() containing x86-64 code) is not portable. We found that some codebases in the dataset were designed and optimized to run on 64-bit platforms. Also, some codebases are platform-dependent and use Linux-specific headers. Such codebases need fundamental changes to their source code before they can be ported to WebAssembly.

\textit{\textbf{Incompatible compiler options.}} Traditional C/C++ compilers provide a set of flags specific to code compilation for native platforms and hardware-specific optimizations. These flags are not available in Emscripten as it targets WebAssembly, which is platform-independent and runs in web environments. For instance, the flag {\useconcmath-march} instructs the compiler to generate code optimized for a user-specified processor by leveraging processor-specific instructions. The optimization flag {\useconcmath-\small{O}fast} is another compiler option available in C/C++ compilers but not in Emscripten. It enables optimizations not permitted by the standard C and C++ specifications, aiming to improve performance potentially at the cost of standards compliance.

Conversely, Emscripten introduces unique compiler options for memory management, multithreading, or exception handling, which differ from those in traditional compilers. Although these default-disabled options often lead to runtime failures rather than build failures, we found some cases where failing to activate Emscripten-specific options at compile time could also cause build errors. We discuss these options in detail in the subsequent sections.

\begin{table}[t]
    \caption{Distribution of challenges leading to build failures.}
  \label{tab:build-results}\small
  \begin{tabular}{lp{8cm}c}
    \toprule
    & \textbf{Build Challenge} & \textbf{Count}\tnote{a} \\
    \midrule
    1 & Undefined symbols & 11 \\
    2 & Missing third-party libraries & 6 \\
    3 & Target-dependent warnings + Werror & 7 \\
    4 & Architecture- and platform-specific code & 5 \\
    5 & Incompatible compiler options & 11 \\
    6 & WebAssembly compiler bugs & 4 \\
    \hline
    \rowcolor{lightgray} & Total build errors & 44 \\
  \bottomrule
\end{tabular}
\end{table}

\textit{\textbf{WebAssembly compiler bugs.}} We observed the build process for a few codebases failed due to bugs in Emscripten. We found 5 new confirmed bugs in Emscripten versions 3.1.54-63, which resulted in build errors across 4 projects in the dataset. Specifically, 4 bugs refer to implementation defects in the LLVM’s {\useconcmath libc}++ headers utilized by Emscripten as the C++ standard library and 1 bug results from the incomplete implementation of functions within the {\useconcmath xlocale.h} header. Moreover, we uncovered 6 other bugs in Emscripten that did not break the compilation process. Instead, they affect the semantics of the Wasm binaries. We discuss these bugs in Section \ref{sec:RQ2-D}.

\section{RQ2: Reliability of Cross-Compiled Code in WebAssembly} \label{sec:RQ2}
\label{sec:RQ2}
To answer RQ2 and specify how well cross-compiling C/C++ code into WebAssembly preserves the code semantics as in native x86-64 binary, we first need a systematic approach to compare the semantics of a single code across two different binaries. Theoretically, two executable codes are called semantically equivalent if they generate the same outputs for every possible input \cite{10.1145/1572272.1572283}. However, such an exhaustive analysis is not feasible in practice. Thus, we have to confine our definition of semantics equivalency between codes. To this end, we resort to a set of test cases provided by developers to dynamically assess the functional behaviors of different components in an open-source codebase.

Given a source program $P$ and a set of tests $T$ evaluating the functional correctness of $P$, we compile both $P$ and $T$ into two different binaries. A test case $t\in T$ might pass or fail when executed in each binary. Let $o_t$ and $o'_t$ be boolean variables taking values corresponding to the outcome of test case $t$ in native binary and WebAssembly, respectively (i.e., equal to 1 if $t$ passes and 0 otherwise). For a program $P$ with a set of tests $T$, we define that $P$ is \textit{semantically equivalent} in native binary and WebAssembly if and only if $\Sigma_{t\in T} |o_t - o'_t| = 0$.

As mentioned in Section \ref{sec:building-codebases}, Emscripten could successfully build the codebases of 81 projects in WebAssembly. Specifically, for these codebases, we enabled the testing option to instruct the build process to compile test cases alongside the source code (e.g., flag {\useconcmath\footnotesize YAML\_CPP\_BUILD\_TESTS} in Fig.~\ref{fig:build}). When we run executable tests for 81 projects in x86-64, 4,687 out of 4,691 tests pass by producing the expected outcomes. Yet, the results surprisingly vary when we run the corresponding JavaScript tests produced by Emscripten\footnote{In this study, we utilized Node.js as the runtime to execute Wasm modules and their corresponding JavaScript glue code.}. Only 4,316 tests pass in WebAssembly, conveying 371 test inconsistencies across two binaries. A deeper analysis shows that a significant portion of WebAssembly test failures are not true alarms for denoting semantics divergence between the two binaries. Rather, they are caused by 3 reasons and are resolvable by modifying compiler settings.

\begin{figure}[t]
    \centering
    \begin{subfigure}[b]{\linewidth}
    \centering
    \begin{tikzpicture}[>=Stealth, node distance=1cm and 1cm, box/.style={draw, minimum height=1cm, align=center}]
    \node (code) at (-0.8,1.5) [align=left, draw, text width=7.9cm, font=\ttfamily, inner xsep=-0mm, inner ysep=-1mm] {
\begin{lstlisting}[style=mystyle2, language=C, xleftmargin=3pt,]
double divideNumbers(double numerator, double denominator) {
    if (denominator == 0) {
        throw std::runtime_error("Division by zero error");
    } return numerator / denominator;
}
int main() {
    double num = 10.0, denom = 0.0;
    try {
        double result = divideNumbers(num, denom);
        std::cout << "Result is " << result << std::endl;
    } catch(const std::runtime_error& e) { // breaks in Wasm
        std::cout << "Caught an exception: " << e.what();
    } return 0;
}
\end{lstlisting}
    };
    
    \node (execution) [right=of code, text width=6.1cm, font=\ttfamily, inner xsep=-2mm, inner ysep=-2mm, yshift=-1cm, xshift=-0.7cm] {
\begin{lstlisting}[style=mystyle2, language=C, xleftmargin=0pt,numbers=none,backgroundcolor=\color{black},basicstyle=\tiny\color{white},commentstyle=\color{green}, frame=single, framesep=0pt]
>_ /home/user/ex.js:128
    throw ex;
    ^
RuntimeError: Aborted(Assertion failed: Exception thrown, but exception catching is not enabled)
at wasm://wasm/0009f852:wasm-function[12]:0xf4c
\end{lstlisting}
    };
    \node[draw, rectangle, minimum height=0.7cm, minimum width=2.2cm,dashed, line width=1pt] (box) at (6.3,2.5) {};
    \node[draw, rectangle, minimum size=0.5cm, fill=custompurple, text=white] (wasm) at (5.9,2.5) {\scriptsize{ex.wasm}};
    \node[draw, rectangle, minimum size=0.5cm, fill=customyellow] (js) at (6.9,2.5) {\scriptsize{ex.js}};
    
    \draw[->] (code) -- (box.west) node[midway, above, font=\scriptsize, right=-1cm,  yshift=-0.5cm] {em++ exception-catching.cpp -o ex.js};

    \draw[->] (js.east) -- (execution.north east) node[midway, right=-0.5cm, font=\scriptsize, sloped, yshift=2mm] {node ex.js};
    \end{tikzpicture}
    \vspace{-0.65em}
    \caption{
    Compiling code by the default settings of Emscripten disables exception handling at runtime. Throwing the exception in line 3 will abort the program without allowing exception catching in line 11 to become executed.}
    \vspace{0.5em}
    \label{fig:exception-catching}
    \end{subfigure}
    \begin{subfigure}[b]{\linewidth}
    \centering
    \begin{tikzpicture}[>=Stealth, node distance=1cm and 1cm, box/.style={draw, minimum height=1cm, align=center}]
    \node (code) at (-0.8,1.5) [align=left, draw, text width=7.9cm, font=\ttfamily, inner xsep=0mm, inner ysep=-1mm, yshift=0.4cm] {
\begin{lstlisting}[style=mystyle2, language=C, xleftmargin=3pt]
typedef void(*func_ptr_void)(const char *);
typedef int(*func_ptr_int)(const char *);
void function(const char *message) {
    printf("This is a message: %s\n", message);
}
int main() {
    func_ptr_void original_func = function;
    func_ptr_int casted_func = (func_ptr_int) original_func;
    original_func("No problem with line 9");
    casted_func("Bomb!"); // breaks in Wasm
}
\end{lstlisting}
    };
    
    \node (execution) [right=of code, text width=6.1cm, font=\ttfamily, inner xsep=-2mm, inner ysep=-2mm, yshift=-0.6cm, xshift=-0.7cm]{
\begin{lstlisting}[style=mystyle2, language=C, xleftmargin=0pt,numbers=none,backgroundcolor=\color{black},basicstyle=\tiny\color{white},commentstyle=\color{green}, frame=single, framesep=0pt]
>_ This is a message: No problem with line 9
/home/user/test/fp.js:128
    throw ex;
    ^
RuntimeError: null function or function signature mismatch
at wasm://wasm/770eb972:wasm-function[4]:0x308
\end{lstlisting}
    };
    \node[draw, rectangle, minimum height=0.7cm, minimum width=2.2cm,dashed, line width=1pt] (box) at (6.3,3.2) {};
    \node[draw, rectangle, minimum size=0.5cm, fill=custompurple, text=white] (wasm) at (5.9,3.2) {\scriptsize\texttt\bfseries{fp.wasm}};
    \node[draw, rectangle, minimum size=0.5cm, fill=customyellow] (js) at (6.9,3.2) {\scriptsize\texttt\bfseries{fp.js}};
    
    \draw[->] (code) -- (box.west) node[midway, above, font=\scriptsize, right=-1cm,  yshift=-0.5cm] {emcc cast-function-pointer.c -o fp.js};

    \draw[->] (js.east) -- (execution.north east) node[midway, right=-0.5cm, font=\scriptsize, yshift=2mm, sloped]{node fp.js};
    \end{tikzpicture}
    \vspace{-0.65em}
    \caption{Type casting of function pointers before calling functions does not work in WebAssembly. The function call in line 10 brings a runtime error in WebAssembly, while it can be executed in native binary without error.}
    \label{fig:function-pointer}
    \end{subfigure}
    \caption{Sample code snippets to show how the default settings of Emscripten produces high-performance WebAssembly code by sacrificing runtime reliability and disabling support for features that bring overhead.}
    \vspace{-0.9em}
    \label{fig:required-flags}
\end{figure}

\subsection{Analysis of Fixable WebAssembly Test Failures}
\label{sec:RQ2-A}
We now discuss the 3 types of WebAssembly test failures that can be resolved by changing compiler settings. We elaborate on the root causes of such test failures and their corresponding solutions.

\textit{\textbf{Default-disabled compiler options.}} First, the default settings of Emscripten are intended to generate high-performance WebAssembly code. It omits support of constructs that introduce additional overhead. For example, supporting exception catching in WebAssembly can lead to performance overhead, as the mechanism required to handle exceptions, such as stack unwinding and state management, is complicated and slows down program execution. Hence, exception catching is disabled in Emscripten by default. In this case, if a program throws an exception, it will not be caught, causing the program to crash (see Fig.~\ref{fig:exception-catching}). To avoid this issue, the compiler flag {\useconcmath\footnotesize NO\_DISABLE\_EXCEPTION\_CATCHING} should be set to enable runtime exception handling.

As another example, in WebAssembly, function pointers/references must be invoked with the same type as declared. This is enforced by the runtime check of type signatures for indirect calls. While calling a function pointer after casting it to a different type works in native binary, it leads to failures in WebAssembly. Instead, one can change such invocations in the code by writing an adapter function that calls the original function without needing to cast. However, this requires familiarity with the codebase to make changes. Emscripten provides the compiler flag {\useconcmath\footnotesize EMULATE\_FUNCTION\_POINTER\_CASTS} as a wrapper to each function in the WebAssembly table\footnote{A data structure used by the WebAssembly module to manage references to functions, allowing indirect function calls.} to fix the parameters and the return type at runtime. Since runtime correction adds overhead, the default settings of Emscripten do not enable this flag, leading to runtime errors in such cases (see Fig.~\ref{fig:function-pointer}).

\textit{\textbf{File access paradigms in C/C++ vs. JavaScript.}} The second issue that causes many tests to fail in WebAssembly comes from different file access paradigms in native C/C++ and JavaScript code. Native C/C++ code usually calls synchronous file APIs in {\useconcmath libc} and {\useconcmath libcxx}, while JavaScript only allows asynchronous file access. Moreover, JavaScript code does not have direct access to the host's file system when running in the sandbox environment provided by a web browser.

Emscripten provides a Virtual File System (VFS) that emulates the local file system, allowing cross-compiled C/C++ code to access files through normal {\useconcmath libc stdio} and {\useconcmath libcxx fstream} APIs \cite{filesystem}. As a limitation, files needed to be accessed by C/C++ code should be preloaded/embedded into the virtual file system when compiling the codebase into WebAssembly. Emscripten does not offer an automated approach for this task. Hence, the user is entrusted with preloading files in the correct path within the virtual file system so that WebAssembly code can access them at runtime.

\textit{\textbf{Memory restrictions.}} 
The third root cause of test failure in WebAssmebly refers to memory restrictions. When running a cross-compiled WebAssembly module, the memory, which handles the execution itself and the data used by the program under execution, involves two segments. One is managed by the WebAssembly module, and the other is under the control of the VM where WebAssembly code is executed. The memory under the control of the VM is commonly referred to as the control stack, which is used to manage the execution flow of the program. It primarily holds the call frames for functions, including the return address and the base frame pointers. For C/C++ code with deeply recursive function calls, the corresponding WebAssembly module obtained by cross-compilation using Emscripten can result in an \textit{exceeded maximum call stack size} error when executed in the Node.js runtime environment. This error indicates the control stack is full and no longer able to store call frames. To reliably cross-compile code with deeply recursive function calls, the WebAssembly code can be optimized using the {\useconcmath-Oz} optimization flag that aggressively minimizes the size of the generated code and reduces the depth of nested calls by doing optimizations such as selective function inlining.

The default settings of Emscripten also impose restrictions on the size of memory under the control of the WebAssembly module, including heap and stack. The default stack size for a WebAssembly module compiled by Emscripten is set to 64KB, which is significantly smaller than that of native binary (e.g., 8MB on Linux and 1MB on Windows). For programs that heavily store data in the stack, cross-compiled Wasm code obtained by Emscripten with default settings brings a \textit{memory access out of bounds} error at runtime. Since there is no way to enlarge the stack size at runtime, one must increase the stack size at compile time using the compiler option {\useconcmath\footnotesize STACK\_SIZE} to make it large enough for the program’s requirements. Otherwise, it silently breaks down the execution while running WebAssembly code. Furthermore, Emscripten restricts the size of heap memory allowed to be allocated by the WebAssembly module. If the module attempts to allocate further than the default size, then a runtime error will occur. Passing the compiler flag {\useconcmath\footnotesize ALLOW\_MEMORY\_GROWTH} to Emscripten at compile time enables the memory array to grow at runtime. These memory limitations also align with Emscripten's goal of generating high-performance WebAssembly code.

\subsection{\sysname: A Differential Testing Framework}\label{sec:RQ2-B}
As discussed, cross-compiling C/C++ programs into Wasm binaries using Emscripten without changing necessary compiler settings, embedding required files in the virtual file system, and relaxing memory restrictions cannot reliably port code semantics. Thus, we need a principled approach to build codebases in WebAssembly so that one ensures all the required settings are applied. Under such circumstances, if at least one test with different outcomes in WebAssembly and x86-64 binaries exists, this will be the notion of code semantics divergence between two binaries. 

Since manually changing compiler settings for each codebase is time-consuming and error-prone, we design and implement an automated approach to build a codebase in WebAssembly. Our approach addresses the build challenges identified in Section \ref{sec:RQ1} as well as the issues leading to unreliable code semantics translation and test failures, as discussed in Section \ref{sec:RQ2-A}. We use this approach to build an integrated framework named \sysname for differential testing between C/C++ programs compiled into native x86-64 binaries and their WebAssembly counterparts. 

\begin{figure}[t]
    \centering
    \vspace{-0.5em}
\includegraphics[width=\linewidth]{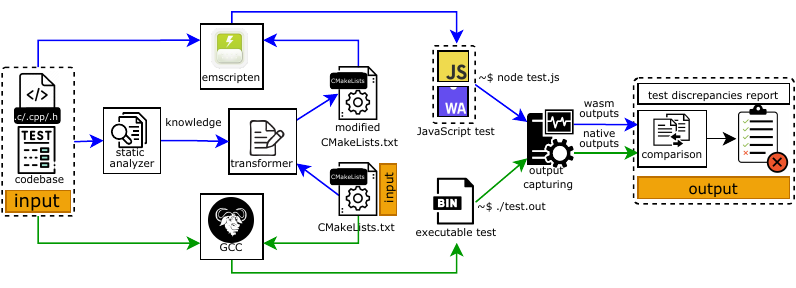}
    \vspace{-2.0em}
    \caption{Architecture of \sysname.}
    \label{fig:wamchecker}
    \vspace{-1.0em}
\end{figure}

\textit{\textbf{\sysname overview.}} The overall workflow of \sysname is illustrated in Fig.~\ref{fig:wamchecker}. Given a codebase, including source code and test cases, \sysname builds the codebase in native binary using the GCC compiler and CMake toolchain (steps highlighted in green in Fig.~\ref{fig:wamchecker}). It then executes executable tests to capture their outcomes.

On the other hand, to build the codebase in WebAssembly (steps highlighted in blue in Fig.~\ref{fig:wamchecker}), \sysname first performs a lightweight static analysis to extract source-level information. For example, code snippets in Fig.~\ref{fig:required-flags} show unexpected behavior at runtime since the two intended flags, {\useconcmath\footnotesize NO\_DISABLE\_EXCEPTION\_CATCHING} and {\useconcmath\footnotesize EMULATE\_FUNCTION\_POINTER\_CASTS}, have not been enabled at compile time. Here, one-step source-level static analysis identifies that these code snippets encompass exception catching and type casting of function pointers. As a result, one not familiar with the codebase's implementation specifics can enable the required flags when compiling into WebAssembly. By leveraging static analysis, \sysname extracts language constructs from the codebase. Then, it passes this knowledge to a transformer that enables required flags based on the code constructs. After modifying compiler settings, \sysname builds the codebase in WebAssembly using Emscripten and subsequently executes JavaScript tests, which instantiate WebAssembly modules, using Node.js to capture test outcomes in WebAssembly.

As the final step, \sysname compares test outputs in two different binaries and reports discrepancies, if any (step marked in black arrow in Fig.~\ref{fig:wamchecker}). The subsequent sections provide detailed descriptions of the static analysis phase and the transformer, which clarifies the entire pipeline. 

\textit{\textbf{Static analyzer.}} \sysname's static analysis component uses CodeQL \cite{4362893}, a code analysis engine. CodeQL converts the codebase into a database format, making it easier to query complex information. It also provides its own declarative query language, similar to SQL, allowing users to write arbitrary queries to specify what patterns they look for. The query Q1 in the following is an example that extracts cast expressions of function pointers and their locations from a codebase:

\noindent
\vspace{-0.2cm}
\renewcommand{\arraystretch}{1.4}
{\fontfamily{put}\selectfont\footnotesize
\[
\text{Q1}
\left\{
\begin{array}{@{}l@{}}
\text{from Cast expr}\\ 
\text{where expr.getUnderlyingType() instanceof FunctionPointerType}\\ 
\text{select expr, expr.getLocation().getFile()}
\end{array}
\right.
\]}
\vspace{-0.1cm}
\renewcommand{\arraystretch}{1}

\noindent
Correspondingly, \sysname's static analyzer uses a set of predefined CodeQL queries to extract codebase constructs. In particular, these queries figure out if each codebase uses multithreading (to set {\useconcmath-pthread}), exception handling (to set {\useconcmath\footnotesize NO\_DISABLE\_EXCEPTION\_CATCHING}), function pointer/reference type casting (to set {\useconcmath\footnotesize EMULATE\_FUNCTION\_POINTER\_CASTS}), and some specific data types, for example, long double data type (to set {\useconcmath\footnotesize PRINTF\_LONG\_DOUBLE}).

To preload the required files in VFS, the static analyzer first extracts constant strings defined in source and test files. Then, it specifies if each string represents a valid absolute or relative path in the local file system on which the compilation takes place. For instance, static analysis identifies three constant strings in the test case in Fig.~\ref{fig:file-packaging}. The first two represent relative paths to files {\useconcmath input.xml} and {\useconcmath test.xml}, respectively. For each file, the static analyzer finds the absolute path to it in the local file system as {\useconcmath src}. It also labels the corresponding path specified in the program as {\useconcmath dst}. Finally, this mapping is passed to the transformer to preload each file from the path {\useconcmath src} in the local file system to the path {\useconcmath dst} in the virtual file system by using {\useconcmath -\thinspace-preload-file src@dst} as a compile-time option.

\begin{figure}[b]
    \centering
\begin{lstlisting}[style=mystyle2, language=C, xleftmargin=8pt, frame=single]
const char* TESTS[] = {"../resources/input/input.xml", "data/test.xml", 0};
for (int i=0; TESTS[i]; i++) {
    XMLDocument doc; const char msg[] = "Stack overflow prevented.";
    doc.LoadFile(TESTS[i]); XMLTest(msg, XML_ELEMENT_DEPTH_EXCEEDED, doc.ErrorID());
}
\end{lstlisting}
    \vspace{-0.5em}
    \caption{Sample test case with file access operations. To preload files in the VFS, \sysname adds compile-time options {\useconcmath-\thinspace-preload-file path\_to\_test@data/test.xml} and {\useconcmath-\thinspace-preload-file path\_to\_input\allowbreak@../resources/input/input.xml} to {\useconcmath\footnotesize CMAKE\_CXX\_FLAGS} by modifying {\useconcmath\footnotesize CMakeLists.txt}.}
    \vspace{-1em}
\label{fig:file-packaging}
\end{figure}
\textit{\textbf{Transformer.}} Based on the source-level information extracted by the static analyzer, \sysname transforms build commands by modifying the CMake build script. For example, if CodeQL outputs a non-empty table for the predefined query Q1, then the transformer enables the compiler flag {\useconcmath\footnotesize EMULATE\_FUNCTION\_POINTER\_CASTS} to ensure that problems around translating function pointers are handled within the compiler's capability.

Besides modifications driven by the outcomes of CodeQL static analysis, the transformer also applies general alterations to build commands. Specifically, it replaces compiler option {\useconcmath-\small{W}error} with {\useconcmath-\small{W}no-error} to avoid build failures that will happen due to target-dependent warnings. Moreover, the transformer removes incompatible compiler options (e.g., {\useconcmath-march}, {\useconcmath-\small{O}fast}) and replaces option {\useconcmath-fstack-protector} with {\useconcmath-fno-stack-protector} to avoid an undefined symbol error caused by enabled SSP. To avoid runtime memory errors, the transformer sets the stack size to 1MB alongside enabling the option {\useconcmath\footnotesize ALLOW\_MEMORY\_GROWTH} and the optimization flag {\useconcmath-\small{O}z}.

\begin{table}[b]
  \caption{\sysname results on 115 projects of the dataset in Table \ref{tab:dataset}.} 
  
  
  \label{tab:results-training}\small
  \begin{tabular}{p{2.2cm} >{\centering}p{2.2cm} >{\centering}p{2.7cm} >{\centering}p{2.8cm} >{\centering\arraybackslash}p{2.1cm}}
    \toprule
    \textbf{Compilation Approach} & \textbf{Compiled Codebases} & \textbf{Total Executable Tests} & \textbf{Codebases With Test Diffs} & \textbf{Total Test Diffs} \\
    \midrule
    Manual & {81} & 4,691 & 58 & 371 \\ 
    \rowcolor{lightgray}\sysname & \textbf{99} (81+18) & 5,638 (4,691+947) & 23 (\textbf{19}+4) & 223 (\textbf{36}+187) \\
  \bottomrule
\end{tabular}
\vspace{-0.5em}
\end{table}

\subsection{Effectiveness of \sysname} \label{sec:RQ2-C}
To evaluate \sysname's effectiveness in fixing build errors, faithfully compiling the codebases into WebAssembly, and precisely identifying discrepancies between code semantics in native and Wasm binaries, we tested it on our collected dataset, containing projects used for identifying initial issues, and on a random sample of additional C/C++ projects distinct from the original dataset.

\textit{\textbf{Experiment 1.}} As discussed in Section \ref{sec:RQ2}, manually compiling 115 codebases in our dataset with the default settings of Emscripten resulted in the successful build for 81 projects with 4,691 executable tests. Such a compilation process also resulted in 371 test discrepancies across 58 codebases when running executable tests in two distinct binaries, x86-64 and WebAssembly. We utilized \sysname across all 115 projects in our dataset. Table \ref{tab:results-training} presents the results of this experiment. \sysname could resolve 25 (57\%) build issues and compile 99 (86\%) codebases, encompassing 5,638 executable tests, into WebAssembly without errors. This included 81 projects (having 4,691 tests) for which the manual compilation was also successful and 18 previously uncompilable ones (having 947 tests). \sysname reported that among 99 codebases, 23 have at least one test producing inconsistent outputs in two binaries. Altogether, it reported 223 test discrepancies across these 23 codebases. Specifically, 36 discrepancies appear in 19 of 81 projects, and 187 arise from 4 of those 18 projects \sysname could build beyond the manual compilation.

Accordingly, \sysname could compile 18 more codebases, showcasing its effectiveness in fixing build issues. Also, in the case of 81 projects, where comparison with manual compilation is possible, these codebases generate $\approx$ 10x fewer test inconsistencies when compiled using \sysname than when manually compiled (36 vs. 371), indicating the higher reliability of Wasm binaries produced using \sysname and its effectiveness in automatically addressing WebAssembly code failures fixable by adjusting compiler settings, as explained in Section \ref{sec:RQ2-A}.

\begin{table}[t]
  \caption{Overview of the collected test dataset.}
  \label{tab:test-dataset}\small
  \begin{tabular}{p{7.1cm}crrr}
    \toprule
    \textbf{Functionality} & \textbf{Projects}\tnote{a} & \textbf{KLOC} & \textbf{Test Cases} & \textbf{Exe. Tests}\tnote{b}\\
    \midrule
    Utility$/$general-purpose (formatting, logging, etc) & 7 & 40\hspace{1mm} & 2,560\hspace{1mm} & 58\hspace{1mm} \\
    Data parsing$/$serialization$/$management & 5 & 212\hspace{1mm} & 501\hspace{1mm} & 141\hspace{1mm} \\
    Mathematical$/$graphical$/$numerical computations & 4 & 46\hspace{1mm} & 186\hspace{1mm} & 5\hspace{1mm} \\
    Algorithms and data structures & 4 & 72\hspace{1mm} & 2,331\hspace{1mm} & 20\hspace{1mm} \\
    \hline
    \rowcolor{lightgray}Total & 20 & 370\hspace{1mm} & 5,578\hspace{1mm} & 224\hspace{1mm} \\
  \bottomrule
\end{tabular}
\vspace{-0.5em}
\end{table}

\textit{\textbf{Experiment 2.}} We also evaluated \sysname on a separate set of unseen codebases to confirm that our results are generalizable beyond the original dataset. To this end, we used the same API-based querying employed in Section \ref{sec:dataset} to randomly collect 35 new C/C++ projects hosted on GitHub, apart from what we used to learn from. We also applied a similar two-phase refinement on the new test dataset, resulting in a final dataset of 20 projects with 5,578 test cases, as summarized in Table \ref{tab:test-dataset}. Table \ref{tab:results-test} demonstrates how effective \sysname is on the test dataset compared to manual compilation. \sysname could build all 20 codebases in WebAssembly, while manual compilation could build 18 projects. Also, running executable tests obtained by manual cross-compilation of 18 codebases resulted in 34 test discrepancies across 14 projects, while it dropped to only 3 test inconsistencies across 2 projects when using \sysname.

We analyzed 226 test discrepancies reported by \sysname in both experiments and identified 220 as true alarms and 6 as false positives. This represents the general effectiveness of \sysname in precisely identifying cases where Emscripten inevitably fails to preserve source code semantics in Wasm binaries. We devote Section \ref{sec:RQ2-D} to elaborating on the root causes of such cases.

\begin{table}[t]
  \caption{\sysname results on 20 projects of the test dataset in Table \ref{tab:test-dataset}. 
  }
  \label{tab:results-test}\small
  \begin{tabular}{p{2.2cm} >{\centering}p{2.2cm} >{\centering}p{2.7cm} >{\centering}p{2.8cm} >{\centering\arraybackslash}p{2.1cm}}
    \toprule
    \textbf{Compilation Approach} & \textbf{Compiled Codebases} & \textbf{Total Executable Tests} & \textbf{Codebases With Test Diffs} & \textbf{Total Test Diffs} \\
    \midrule
    Manual & 18 & 221 & 14 & 34 \\
    \rowcolor{lightgray}\sysname & \textbf{20} (18+2) & 224 (221 + 3) & 2 (\textbf{2}+0) & 3 (\textbf{3}+0) \\
  \bottomrule
\end{tabular}
\vspace{-0.5em}
\end{table}

\textit{\textbf{Limitations.}} The main limitation of \sysname, which leads to a few false alarms, is the heuristic-based method for automatic file embedding. Currently, \sysname does not use advanced types of analysis (e.g., string analysis \cite{10.1145/2786805.2786879}) to map paths in the local file system to the ones in the virtual file system. The correct path to a file in the VFS may not be defined within the code as a constant string (e.g., a case where the path is obtained by concatenating two strings).

In addition, \sysname does not perform source code adaption. Instead, it focuses on addressing build failures that are resolvable by modifying compiler settings and build commands. Among 6 categories of build challenges, \sysname can fully address build failures arising from incompatible compiler options and target-dependent warnings. Furthermore, it can partially resolve issues coming from undefined symbol errors (limited to errors resulting from enabled SSP or symbols defined within libraries that are supported in case of enabling specific compiler options) and missing third-party libraries (confined to libraries available on \cite{emports}). However, \sysname cannot address failures occurring due to architecture- or platform-specific instructions within the source code and those stemming from unsupported API calls. Addressing these issues requires changes in the source code, which falls outside the scope of this paper but would be a valuable direction for future work.

\subsection{Analysis of Semantics Divergence Among Binaries}
\label{sec:RQ2-D}
We analyzed the root causes of true test discrepancies reported by \sysname and categorized them into 4 categories, which are discussed in this section.

\begin{figure}[t]
    \begin{subfigure}[b]{\linewidth}
\begin{lstlisting}[style=mystyle2, language=C, xleftmargin=8pt, frame=single]
std::unordered_map<vertex_id_t, int> welsh_powell_coloring(const GRAPH& graph) {
    using degree_vertex_pair = std::pair<int, vertex_id_t>;
    // Step 1: Sort vertices by degree in descending order
    std::vector<degree_vertex_pair> degree_vertex_pairs;
    for (const auto& [vertex_id, _] : graph.get_vertices()) {
        int degree = properties::vertex_degree(graph, vertex_id);
        degree_vertex_pairs.emplace_back(degree, vertex_id);
    } std::sort(degree_vertex_pairs.rbegin(), degree_vertex_pairs.rend());
    // Step 2: Assign colors to vertices
    std::unordered_map<vertex_id_t, int> color_map;
    for (const auto [_, curr_vertex] : degree_vertex_pairs) {
        int color = 0; // Start with color 0 & check colors of adjacent vertices
        for (const auto& neighbor : graph.get_neighbors(curr_vertex)) {
            // If neighbor is already colored with this color, increment the color
            if (color_map.contains(neighbor) && color_map[neighbor] == color)
                color = color + 1;
        } color_map[curr_vertex] = color; // Assign the color to the curr vertex
    } return color_map;
}
\end{lstlisting}
    \vspace{-0.25em}
    \caption{Implementation of Welsh Powell graph coloring algorithm using Graaf library \cite{graaf}.}
    \label{fig:welsh-coloring-algorithm}
    \end{subfigure}
    \vfill 
    \vspace{0.5em}
    \begin{subfigure}[b]{\linewidth}
    \begin{lstlisting}[style=mystyle2, language=C, xleftmargin=8pt, frame=single]
TYPED_TEST(WelshPowellTest, CompleteGraph) {
    using graph_t = typename TestFixture::graph_t; graph_t graph{};
    std::vector<vertex_id_t> vertices; // Graph vertices
    for (int i = 0; i < 4; ++i) {
        vertices.push_back(graph.add_vertex(i));
    } for (size_t i = 0; i < vertices.size(); i++) { // Add edges
        for (size_t j = i + 1; j < vertices.size(); j++) {
            graph.add_edge(vertices[i], vertices[j], 1);
        }
    } auto coloring = welsh_powell_coloring(graph);
    std::unordered_map<vertex_id_t, int> expected = {{3,0},{2,1},{1,2}{0,3}};
    ASSERT_EQ(coloring, expected); // Check if the obtained coloring matches the expected coloring
}
\end{lstlisting}
    \vspace{-0.25em}
    \caption{Test case for evaluating the correctness of the algorithm in Fig. \ref{fig:welsh-coloring-algorithm} on a complete graph with four vertices.}
    \label{fig:welsh-coloring-test}
    \end{subfigure}
    \caption{Example to showcase how various implementations of data structures across different standard libraries affect the program's flow and the test case output.}
    \vspace{-1.0em}
    \label{fig:welsh-coloring}
\end{figure}

\textit{\textbf{Different standard libraries.}} When compiling code into WebAssembly, Emscripten employs {\useconcmath musl} and LLVM's {\useconcmath libc}++ as the C and C++ standard libraries, respectively. However, when compiling into native x86-64 binary, the compiler utilizes the operating system's native standard libraries, in our case, {\useconcmath glibc} and {\useconcmath libstdc}++. Various implementations of standard libraries can result in semantics discrepancies between WebAssembly and native binaries. The C standard library {\useconcmath musl} is lightweight with unique implementations of functions that can lead to subtle differences in memory management, thread handling, and locale support compared to {\useconcmath glibc}. Similarly, {\useconcmath libc}++ handles certain data structures and exceptions differently from {\useconcmath libstdc}++. Thus, the code compiled into WebAssembly might exhibit variations in functional behavior compared to its native x86-64 counterpart, which could affect the application's logic.

\begin{figure}[t]
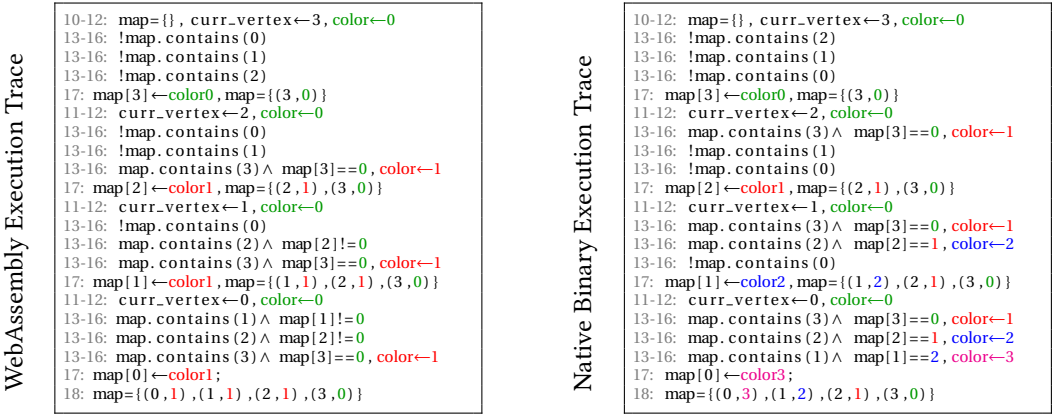

\begin{minipage}[b]{0.05\linewidth} 
    \raisebox{0.1\height}{\rotatebox{90}{WebAssembly Execution Trace}} 
\end{minipage}
\begin{subfigure}[b]{0.40\linewidth}
\begin{lstlisting}[style=mystyle2, language=C, xleftmargin=0pt, numbers=none, frame=single]
@{\color{gray}10-12:}@ @{\thinspace}@map={},@{\thinspace}@curr_vertex@{$\leftarrow$}@3,@{\thinspace}@@{\color{codegreen}color$\leftarrow$0}@
@{\color{gray}13-16:}@ !map.contains(0)
@{\color{gray}13-16:}@ !map.contains(1)
@{\color{gray}13-16:}@ !map.contains(2)
@{\color{gray}17:}@ map[3]@{$\leftarrow$}@@{\color{codegreen}color0}@,@{\thinspace}@map={(3,@{\color{codegreen}0}@)}
@{\color{gray}11-12:}@ curr_vertex@{$\leftarrow$}@2,@{\thinspace}@@{\color{codegreen}color$\leftarrow$0}@
@{\color{gray}13-16:}@ !map.contains(0)
@{\color{gray}13-16:}@ !map.contains(1)
@{\color{gray}13-16:}@ @{\thinspace}@map.contains(3)@{$\land$}@ map[3]==@{\color{codegreen}0}@,@{\thinspace}@@{\color{red}color$\leftarrow$1}@
@{\color{gray}17:}@ map[2]@{$\leftarrow$}@@{\color{red}color1}@,@{\thinspace}@map={(2,@{\color{red}1}@),(3,@{\color{codegreen}0}@)} 
@{\color{gray}11-12:}@ curr_vertex@{$\leftarrow$}@1,@{\thinspace}@@{\color{codegreen}color$\leftarrow$0}@
@{\color{gray}13-16:}@ !map.contains(0)
@{\color{gray}13-16:}@ @{\thinspace}@map.contains(2)@{$\land$}@ map[2]!=@{\color{codegreen}0}@
@{\color{gray}13-16:}@ @{\thinspace}@map.contains(3)@{$\land$}@ map[3]==@{\color{codegreen}0}@,@{\thinspace}@@{\color{red}color$\leftarrow$1}@
@{\color{gray}17:}@ map[1]@{$\leftarrow$}@@{\color{red}color1}@,@{\thinspace}@map={(1,@{\color{red}1}@),(2,@{\color{red}1}@),(3,@{\color{codegreen}0}@)}
@{\color{gray}11-12:}@ curr_vertex@{$\leftarrow$}@0,@{\thinspace}@@{\color{codegreen}color$\leftarrow$0}@
@{\color{gray}13-16:}@ map.contains(1)@{$\land$}@ map[1]!=@{\color{codegreen}0}@
@{\color{gray}13-16:}@ map.contains(2)@{$\land$}@ map[2]!=@{\color{codegreen}0}@
@{\color{gray}13-16:}@ map.contains(3)@{$\land$}@ map[3]==@{\color{codegreen}0}@,@{\thinspace}@@{\color{red}color$\leftarrow$1}@
@{\color{gray}17:}@ map[0]@{$\leftarrow$}@@{\color{red}color1}@;
@{\color{gray}18:}@ map={(0,@{\color{red}1}@),(1,@{\color{red}1}@),(2,@{\color{red}1}@),(3,@{\color{codegreen}0}@)}
\end{lstlisting}
\end{subfigure}
\hfill 
\begin{minipage}[b]{0.05\linewidth} 
    \raisebox{0.1\height}{\rotatebox{90}{Native Binary Execution Trace}} 
\end{minipage}
\begin{subfigure}[b]{0.40\linewidth}
\begin{lstlisting}[style=mystyle2, language=C, xleftmargin=0pt, numbers=none, frame=single]
@{\color{gray}10-12:}@ @{\thinspace}@map={},@{\thinspace}@curr_vertex@{$\leftarrow$}@3,@{\thinspace}@@{\color{codegreen}color$\leftarrow$0}@
@{\color{gray}13-16:}@ !map.contains(2)
@{\color{gray}13-16:}@ !map.contains(1)
@{\color{gray}13-16:}@ !map.contains(0)
@{\color{gray}17:}@ @{\thinspace}@map[3]@{$\leftarrow$}@@{\color{codegreen}color0}@,@{\thinspace}@map={(3,@{\color{codegreen}0}@)}
@{\color{gray}11-12:}@ curr_vertex@{$\leftarrow$}@2,@{\thinspace}@@{\color{codegreen}color$\leftarrow$0}@
@{\color{gray}13-16:}@ @{\thinspace}@map.contains(3)@{$\land$}@ @{\thinspace}@map[3]==@{\color{codegreen}0}@,@{\thinspace}@@{\color{red}color$\leftarrow$1}@
@{\color{gray}13-16:}@ !map.contains(1)
@{\color{gray}13-16:}@ !map.contains(0)
@{\color{gray}17:}@ @{\thinspace}@map[2]@{$\leftarrow$}@@{\color{red}color1}@,@{\thinspace}@map={(2,@{\color{red}1}@),(3,@{\color{codegreen}0}@)}
@{\color{gray}11-12:}@ curr_vertex@{$\leftarrow$}@1,@{\thinspace}@@{\color{codegreen}color$\leftarrow$0}@
@{\color{gray}13-16:}@ @{\thinspace}@map.contains(3)@{$\land$}@ @{\thinspace}@map[3]==@{\color{codegreen}0}@,@{\thinspace}@@{\color{red}color$\leftarrow$1}@
@{\color{gray}13-16:}@ @{\thinspace}@map.contains(2)@{$\land$}@ @{\thinspace}@map[2]==@{\color{red}1}@,@{\thinspace}@@{\color{blue}color$\leftarrow$2}@
@{\color{gray}13-16:}@ !map.contains(0)
@{\color{gray}17:}@ @{\thinspace}@map[1]@{$\leftarrow$}@@{\color{blue}color2}@,@{\thinspace}@map={(1,@{\color{blue}2}@),(2,@{\color{red}1}@),(3,@{\color{codegreen}0}@)}
@{\color{gray}11-12:}@ curr_vertex@{$\leftarrow$}@0,@{\thinspace}@@{\color{codegreen}color$\leftarrow$0}@
@{\color{gray}13-16:}@ @{\thinspace}@map.contains(3)@{$\land$}@ @{\thinspace}@map[3]==@{\color{codegreen}0}@,@{\thinspace}@@{\color{red}color$\leftarrow$1}@
@{\color{gray}13-16:}@ @{\thinspace}@map.contains(2)@{$\land$}@ @{\thinspace}@map[2]==@{\color{red}1}@,@{\thinspace}@@{\color{blue}color$\leftarrow$2}@
@{\color{gray}13-16:}@ @{\thinspace}@map.contains(1)@{$\land$}@ @{\thinspace}@map[1]==@{\color{blue}2}@,@{\thinspace}@@{\color{magenta}color$\leftarrow$3}@
@{\color{gray}17:}@ @{\thinspace}@map[0]@{$\leftarrow$}@@{\color{magenta}color3}@;
@{\color{gray}18:}@ @{\thinspace}@map={(0,@{\color{magenta}3}@),(1,@{\color{blue}2}@),(2,@{\color{red}1}@),(3,@{\color{codegreen}0}@)}
\end{lstlisting}
\end{subfigure}
\vspace{-0.5em}
\caption{Execution traces for the algorithm and the test case of Fig. \ref{fig:welsh-coloring} in Wasm and x86-64 binaries.}
\label{fig:execution-trace}
\vspace{-0.5em}
\end{figure}

Taking the implementation of the Welsh Powell graph coloring algorithm shown in Fig.~\ref{fig:welsh-coloring-algorithm} as an example, we illustrate how different implementations of standard libraries can affect the program's logic. For the given test case in Fig.~\ref{fig:welsh-coloring-test}, which examines the correctness of the algorithm on a complete graph with four vertices, the expected coloring assigns four different colors (from 0 to 3) to four vertices. This algorithm operates as expected in native binary since the method {\useconcmath get\_neighbors} in line 13 returns the neighbors of the {\useconcmath curr\_vertex} in descending order (i.e., in each loop iteration, the code checks the color of a neighbor with higher {\useconcmath vertex\_id} before the ones with lower {\useconcmath vertex\_id}s). When compiling this code into WebAssembly, the set returned by {\useconcmath get\_neighbors} is different from what is in the native binary. In this case, the {\useconcmath for} loop in line 13 first checks neighbors with lower {\useconcmath vertex\_id}s. Fig.~\ref{fig:execution-trace} demonstrates how the different order of neighbors checked in each iteration affects the test outcome in native and WebAssembly binaries.

As another example, the test case in Fig.~\ref{fig:easylogging}, taken from the logging library Easylogging \cite{easylogging}, has been designed to ensure the logging process behaves as expected in case of an error in file access operations. First, the test attempts to open a file that does not exist in the file system. If the file unexpectedly opens, the code triggers a failure by calling {\useconcmath\footnotesize FAIL()}. Otherwise, the code logs a message using the logging macro {\useconcmath\footnotesize PLOG(INFO)}. The test anticipates the recorded log to be a string including the current date besides the message {\useconcmath"This is plog: No such file or directory"} followed by a specific error number (errno) {\useconcmath\small"2"} corresponding to {\useconcmath\footnotesize ENOENT} macro defined at {\useconcmath errno.h}, a standard library in {\useconcmath libc}. However, since the C standard library in Emscripten adheres to {\useconcmath musl} provided by the WebAssembly System Interface (WASI), it defines a different errno {\useconcmath\small"44"} for an error that occurs when attempting to open a non-existent file. Thus, the log statement retrieved by {\useconcmath tail(1)} does not match the string {\useconcmath expected}, and the test fails in WebAssembly.

\begin{figure}[t]
    \centering
\begin{lstlisting}[style=mystyle2, language=C, xleftmargin=8pt, frame=single]
TEST(PLogTest, WriteLog) {
    std::fstream file("/a/file/that/does/not/exist.txt", std::fstream::in);
    if (file.is_open()) { FAIL(); /* we do not expect to open  file */ }
    PLOG(INFO) << "This is plog";
    std::string expected = BUILD_STR(getDate() << " This is plog: No such file or directory [2]\n");
    std::string actual = tail(1);
    EXPECT_EQ(expected, actual); // Check if the log is as expected
}
\end{lstlisting}
    \vspace{-0.5em}
    \caption{Test case with different outputs in two binaries due to inconsistent error numbers defined by {\useconcmath errno.h}.}
    \vspace{-0.5em}
\label{fig:easylogging}
\end{figure}

Overall, hashing mechanisms employed by different standard libraries for data structures (e.g., set, map) affect the order of elements when insertion. In addition, error numbers and static members of template classes such as {\useconcmath std::numeric\_limits}, which represents properties of arithmetic types, may get various values across different libraries. We found 13 discrepancies among test outputs, which stem from different implementations of standard libraries used by GCC and Emscripten.

\textit{\textbf{Unsupported system calls and APIs.}} Native binary is directly executed by the operating system on the hardware, taking full advantage of the processor and system features available on a platform. On the other hand, WebAssembly code is executed within a sandboxed virtual machine that restricts resource access (e.g., file system, network interfaces). WebAssembly relies on the browser APIs or the embedding environment for file access, networking, and other system-level operations. As mentioned in the previous sections, a class of system-level POSIX APIs is unportable through Emscripten into WebAssembly. If a code uses such APIs, Emscripten generates an undefined symbol error when compiling that code. There are unimplemented system calls (e.g., {\useconcmath fork}, {\useconcmath exec}) that do not break the compilation process by introducing errors. In contrast, they always fail when executed. For example, the code snippet in Fig.~\ref{fig:fork} uses the system call {\useconcmath fork}. When compiling this code into WebAssembly using Emscripten, the compilation process does not introduce any errors. However, when running the cross-compiled WebAssembly code, the conditions {\useconcmath if} in line 3 are not evaluated as true. Since Emscripten does not emulate the functionality of {\useconcmath fork}, calling this system call in line 2 of the function {\useconcmath getChild} returns -1. As another example, given a C code that implements TCP networking using POSIX socket APIs, Emscripten tries to mimic the connection over WebSockets. Yet, this emulation is currently incomplete, which causes runtime failures of unsupported socket primitives. We identified 197 discrepancies among the test outputs, which occurred due to unsupported system calls/APIs.

\begin{figure}[t]
    \centering
    \begin{subfigure}[b]{0.45\linewidth}
    \centering
\begin{lstlisting}[style=mystyle2, language=C, xleftmargin=8pt, frame=single]
int getChild(){
    int pid = fork();
    if (pid > 0) { return pid; }
    return 0;
}
\end{lstlisting}
\end{subfigure}
    \hfill
    \begin{subfigure}[b]{0.52\linewidth}
    \centering
\begin{lstlisting}[style=mystyle2, language=C, xleftmargin=8pt, frame=single]
extern int getChild();
int child_pid = getChild();
if (child_pid > 0) {
    /* some lines of code */ 
} 
\end{lstlisting}
\end{subfigure}
    \vspace{-0.5em}
    \caption{A code snippet calling system call {\useconcmath fork}.}
    \label{fig:fork}
    \vspace{-0.5em}
\end{figure}
\begin{figure}[t]
    \centering
    \begin{subfigure}[b]{0.45\linewidth}
    \centering
\begin{lstlisting}[style=mystyle2, language=C, xleftmargin=8pt, frame=single]
int helloWorld();
int main() { helloWorld(); }
\end{lstlisting}
\end{subfigure}
    \hfill
    \begin{subfigure}[b]{0.52\linewidth}
    \centering
\begin{lstlisting}[style=mystyle2, language=C, xleftmargin=8pt, frame=single]
// Inconsistent function definition
void helloWorld() { printf("hello world!\n"); }
\end{lstlisting}
\end{subfigure}
    \vspace{-0.5em}
    \caption{Example of inconsistent function declaration and definition in C.}
    \label{fig:function_mismatch}
    \vspace{-0.5em}
\end{figure}

\textit{\textbf{WebAssembly language features.}}
The core specification of WebAssembly is intended to address the problem of safe, fast, portable, low-level code on the web \cite{WebAssemblyPLDI}. To achieve this goal, WebAssembly introduces new features enforcing differences in code execution compared to native binaries. For example, in WebAssembly, both direct and indirect calls are subject to dynamic type checking, which prevents the execution of functions in case of a signature mismatch. Accordingly, when the sample code in Fig.~\ref{fig:function_mismatch} is compiled and run as a Wasm binary, it generates an \textit{unreachable} runtime error arising from the function signature mismatch between the declaration and definition of the function {\useconcmath helloWorld}. Yet, this code can be successfully compiled and run as a x86-64 binary. We figured out that 1 test shows inconsistent behavior in two binaries due to runtime type checks.

\begin{figure}[b]
    \centering
    \begin{subfigure}[b]{0.65\linewidth}
    \centering
\begin{lstlisting}[style=mystyle2, language=C, xleftmargin=8pt, xrightmargin=0pt, frame=single]
dirent* _entry;
DIR* _dir = ::opendir("/path/to/dir");
if(!_dir) {std::cout << "dir is null"; exit(0);}
while((_entry = ::readdir(_dir)) != nullptr) {
    if (!strcmp(_entry->d_name, "emp")) {
        /* some lines of code */
    }
}
\end{lstlisting}
    \end{subfigure}
    \hfill
    \begin{subfigure}[b]{0.32\linewidth}
\begin{lstlisting}[style=mystyle2, language=C, xleftmargin=0pt,numbers=none,backgroundcolor=\color{black},basicstyle=\ttfamily\tiny\color{white},frame=single]
 >_:/path/to/dir$ tree
 .
 |-- data.txt
 |-- emp
 |-- non-emp
     `-- program.c

 2 directories, 2 files
\end{lstlisting}
    \end{subfigure}
    \vspace{-0.5em}
    \caption{Example to show the effect of a bug in the Emscripten's VFS emulation on the program output.}
    \label{fig:empty-dir}
    \vspace{-0.5em}
\end{figure}

\textit{\textbf{Compiler bugs.}} 
Besides using different standard libraries, unsupported API calls, and WebAssembly's unique features, we found implementation bugs in Emscripten that are responsible for unexpected Wasm code behaviors. In these cases, although the WebAssembly code was successfully built, discrepancies were introduced in their semantics, which resulted in different test outcomes in WebAssembly and x86-64 binaries. For example, an implementation bug in file system emulation prevents empty directories from being loaded into the virtual file system when preloading. Fig.~\ref{fig:empty-dir} shows how this bug can affect the program's outcome or even change the program's flow. The right-side figure visualizes the file system's structure starting from a directory {\useconcmath dir}. This directory includes file {\useconcmath data.txt} and two subdirectories, {\useconcmath emp} and {\useconcmath non-emp}, which are empty and contain a single file {\useconcmath program.c}, respectively. The code snippet on the left opens the directory {\useconcmath dir} and iterates over the entries within it. It also executes the body of {\useconcmath if} in line 5 when encountering the subdirectory {\useconcmath emp}. Compiling and running this code as a native binary results in the code execution at line 6. However, when we use Emscripten with the compiler option {\useconcmath -\thinspace-preload-file} to preload the entire directory {\useconcmath dir} in the VFS, the empty directory {\useconcmath emp} is excluded. Thus, cross-compiled Wasm code does not execute the {\useconcmath if} block. We located 7 bugs in Emscripten, including 6 new confirmed ones, which resulted in 14 test inconsistencies across the two binaries.

\section{Discussion}
This section summarizes the major findings of our study and discuss its potential validity threats.

\textit{\textbf{Study findings.}} We hope these findings highlight opportunities to improve Wasm compilers and streamline high-level language code reuse in widespread WebAssembly applications.

\begin{itemize}
[leftmargin=*, itemindent=9pt]
    \item The current version of WebAssembly imposes a tradeoff between code performance and code reusability. Although recent proposals extend WebAssembly's capabilities to support various C/C++ language features, such as exception handling, these proposals need to be improved in terms of performance cost.
    \item Porting legacy C/C++ code to WebAssembly often has challenges, some of which are resolvable by modifying compiler settings when cross-compiling code (e.g., removing troublesome compiler options), and some others require significant changes in the source code to make it portable (e.g., adapting unsupported API calls and architecture-specific instructions).
    \item Though compilers are usually assumed to be trusted software, the default settings of Emscripten as a widely-used WebAssembly compiler do not reliably cross-compile code semantics due to default-disabled compiler options, different file access paradigms in WebAssembly compared to native binary, and memory limitations imposed when generating Wasm code. 
    \item Since WebAssembly code runs in a sandbox environment without access to underlying system resources, file system access operations must be emulated to perform through a virtual file system over browser APIs. Constructing a well-structured virtual file system adds complexity to the code compilation process (e.g., necessary file embeddings at compile time).
    \item Emscripten compiler, even with user-specified settings, does not necessarily port the exact code semantics due to several reasons: (1) Traditional C/C++ compilers and WebAssembly compilers use various standard libraries, which may implement data structures and algorithms differently. (2) WebAssembly compilers do not fully emulate the functionality of all system-level APIs and system calls. (3) WebAssembly enforces new features (e.g., dynamic check of function signatures) to make the code execution safe and fast, which may differ from native binaries. (4) Despite recent advancements, Wasm compilers still contain bugs, which affect the code semantics translation.
\end{itemize}

\textit{\textbf{Threats to validity.}} Our study might be potentially subject to several threats, including the representativeness of the chosen codebases and the correctness of the analysis methodology. Regarding the representativeness of the codebases in the dataset, we selected codebases of real-world projects hosted on GitHub containing test cases and {\useconcmath CMakeLists.txt}. There might exist build challenges not covered by this study due to the criteria specified for the data collection process. 

Another threat concerns the correctness of the analysis methodology. We confine the analysis of code semantics to those functional behaviors that can be evaluated using a set of predefined test cases. To alleviate this threat, we used real-world codebases from public repositories with over 50 stars to ensure the diversity and comprehensiveness of the test cases.

\section{Related Work}
\label{sec:RelatedWork}
\textit{\textbf{Analyzing binary security of WebAssembly.}}
Several papers discuss the security behavior of code cross-compiled into WebAssembly \cite{stievenart2022security, 255318, 9724846, WasmBinRealWorld}. The closest work to ours is by Sti{\'e}venart et al. \cite{stievenart2022security}, which studies the behavior of vulnerable codes compiled into WebAssembly compared to their x86-64 counterparts. This work focuses on a group of discrepancies critical from a security point of view to warn developers that compiling code into Wasm binary may expose the security of WebAssembly applications to risk. In \cite{stievenart2022security}, the authors manually compiled Juliet C/C++ Test Suite \cite{6329885}, a dataset of small, synthetic, vulnerable programs divided into several classes based on CWE-ID, into WebAssembly. Though the syntactic structures of programs in a specific class are different, they have no semantics distinction. Instead of targeting synthetic, vulnerable code, we test real-world programs using an automated framework designed for differential testing between two binaries by taking advantage of open-source tests.

\textit{\textbf{Differential testing for bug diagnosis in compilers.}}
An effective technique for detecting bugs in software is differential testing, which involves comparing multiple implementations of the same specification \cite{McKeeman1998DifferentialTF}. Specifically, some studies use differential testing to find bugs in compilers \cite{yang2011finding, RustSmith, LeRe, CompDiff, RandIR}. Csmith \cite{yang2011finding} is a randomized test case generation tool that leverages differential testing for bug discovery in C compilers. It utilizes a specific grammar to randomly generate idiomatic code with a top-level function main that calculates a checksum to convey the program’s result. This way, Csmith compares checksum outputs among different C compilers to report potential bugs. Inspired by Csmith, RustSmith \cite{RustSmith} has been introduced as a randomized program generator for differential testing of Rust compilers. It generates programs that conform to the advanced type system of Rust in terms of rules related to borrowing and lifetimes and that are guaranteed to yield a well-defined output. These approaches use differential testing to detect bugs in compilers of a specific language. In contrast, we use this technique to identify cases where C/C++ compilers and those for WebAssembly translate the code semantics differently.

\textit{\textbf{Differential testing among WebAssembly runtimes.}}
 Researchers have recently used differential testing to analyze discrepancies among Wasm runtimes \cite{perenyi2020stack, hamidy2020differential, WADIFF, Warpdiff, WRTester}. Since WebAssembly does not mandate a specific implementation, WebAssembly engines may vary in how they apply the specification. Per{\'e}nyi et al. \cite{perenyi2020stack} and Hamidy et al. \cite{hamidy2020differential} employed differential fuzz testing to identify bugs and discrepancies in browser-based WebAssembly engines. WADIFF \cite{WADIFF} generates sufficient meaningful test cases using a DSL transformer, which converts the semantics of WebAssembly specification into structured DSL, combined with a symbolic execution engine for test case generation. WRTester \cite{WRTester} creates WebAssembly code by assembling basic elements from real-world Wasm binaries and applies mutation strategies to diversify the code. WarpDiff \cite{Warpdiff} leverages differential testing to detect performance issues in server-side Wasm runtimes. In normal cases, the execution time of the same test case on different runtimes should follow a stable oracle ratio, reflecting the systematic performance gaps between various runtimes. WarpDiff identifies abnormal cases where the execution time ratio deviates from the oracle ratio.
\section{Conclusion}
\label{sec:Conclusion}
We conducted a two-phase study to investigate the capability of WebAssembly compilers to realize code reusability. Specifically, we categorized failures one might encounter when compiling legacy C/C++ codebases into WebAssembly. We also developed a differential testing framework, \sysname, to compare semantics equivalency between Wasm and native x86-64 binaries. Using \sysname and a set of open-source C/C++ projects, we showed that WebAssembly compilers do not necessarily preserve source code semantics in the resulting Wasm binaries.
Our dataset of open-source codebases, \sysname's source code, and the scripts for reproducing the results are available at \href{https://github.com/SaraBaradaran/WasmChecker}{https://github.com/SaraBaradaran/WasmChecker}.
\bibliography{acmart}


\begin{thebibliography}{55}


\ifx \showCODEN    \undefined \def \showCODEN     #1{\unskip}     \fi
\ifx \showDOI      \undefined \def \showDOI       #1{#1}\fi
\ifx \showISBNx    \undefined \def \showISBNx     #1{\unskip}     \fi
\ifx \showISBNxiii \undefined \def \showISBNxiii  #1{\unskip}     \fi
\ifx \showISSN     \undefined \def \showISSN      #1{\unskip}     \fi
\ifx \showLCCN     \undefined \def \showLCCN      #1{\unskip}     \fi
\ifx \shownote     \undefined \def \shownote      #1{#1}          \fi
\ifx \showarticletitle \undefined \def \showarticletitle #1{#1}   \fi
\ifx \showURL      \undefined \def \showURL       {\relax}        \fi
\providecommand\bibfield[2]{#2}
\providecommand\bibinfo[2]{#2}
\providecommand\natexlab[1]{#1}
\providecommand\showeprint[2][]{arXiv:#2}

\bibitem[gcc(1987)]%
        {gcc}
 \bibinfo{year}{1987}\natexlab{}.
\newblock \bibinfo{booktitle}{\emph{{GCC, the GNU Compiler Collection}}}.
\newblock
\urldef\tempurl%
\url{https://gcc.gnu.org}
\showURL{%
\tempurl}


\bibitem[boo(2000)]%
        {boost}
 \bibinfo{year}{2000}\natexlab{}.
\newblock \bibinfo{booktitle}{\emph{{Boost C++ Libraries}}}.
\newblock
\urldef\tempurl%
\url{https://github.com/boostorg/boost}
\showURL{%
\tempurl}


\bibitem[can(2008)]%
        {can_i_use_wasm}
 \bibinfo{year}{2008}\natexlab{}.
\newblock \bibinfo{title}{{Can I use: up-to-date browser support tables for support of front-end web technologies on desktop and mobile web browsers}}.
\newblock
\newblock
\urldef\tempurl%
\url{https://caniuse.com/wasm}
\showURL{%
\tempurl}


\bibitem[Gte(2008)]%
        {Gtest}
 \bibinfo{year}{2008}\natexlab{}.
\newblock \bibinfo{booktitle}{\emph{{GoogleTest}}}.
\newblock
\urldef\tempurl%
\url{https://github.com/google/googletest}
\showURL{%
\tempurl}


\bibitem[yam(2008)]%
        {yaml-cpp}
 \bibinfo{year}{2008}\natexlab{}.
\newblock \bibinfo{booktitle}{\emph{{yaml-cpp: A YAML parser and emitter in C++}}}.
\newblock
\urldef\tempurl%
\url{https://github.com/jbeder/yaml-cpp}
\showURL{%
\tempurl}


\bibitem[Cat(2010)]%
        {Catch2}
 \bibinfo{year}{2010}\natexlab{}.
\newblock \bibinfo{booktitle}{\emph{{Catch2}}}.
\newblock
\urldef\tempurl%
\url{https://github.com/catchorg/Catch2}
\showURL{%
\tempurl}


\bibitem[ems(2010)]%
        {emscripten}
 \bibinfo{year}{2010}\natexlab{}.
\newblock \bibinfo{booktitle}{\emph{{Emscripten: a complete open source LLVM-based compiler toolchain for WebAssembly}}}.
\newblock
\urldef\tempurl%
\url{https://emscripten.org}
\showURL{%
\tempurl}


\bibitem[eas(2012)]%
        {easylogging}
 \bibinfo{year}{2012}\natexlab{}.
\newblock \bibinfo{booktitle}{\emph{{Easylogging++: A logging library for C++ applications}}}.
\newblock
\urldef\tempurl%
\url{https://github.com/abumq/easyloggingpp}
\showURL{%
\tempurl}


\bibitem[git(2013)]%
        {githubapi}
 \bibinfo{year}{2013}\natexlab{}.
\newblock \bibinfo{title}{{GitHub REST API}}.
\newblock
\newblock
\urldef\tempurl%
\url{https://docs.github.com/en/rest/search/search?apiVersion=2022-11-28}
\showURL{%
\tempurl}


\bibitem[emp(2014)]%
        {emports}
 \bibinfo{year}{2014}\natexlab{}.
\newblock \bibinfo{booktitle}{\emph{{Emscripten ported libraries}}}.
\newblock
\urldef\tempurl%
\url{https://github.com/emscripten-ports}
\showURL{%
\tempurl}


\bibitem[bin(2015)]%
        {binaryen}
 \bibinfo{year}{2015}\natexlab{}.
\newblock \bibinfo{booktitle}{\emph{{Binaryen: an optimizer and toolchain library for WebAssembly}}}.
\newblock
\urldef\tempurl%
\url{https://github.com/WebAssembly/binaryen}
\showURL{%
\tempurl}


\bibitem[Ass(2017)]%
        {AssemblyScript}
 \bibinfo{year}{2017}\natexlab{}.
\newblock \bibinfo{booktitle}{\emph{{AssemblyScript: a TypeScript-like language for WebAssembly}}}.
\newblock
\urldef\tempurl%
\url{https://github.com/AssemblyScript/assemblyscript}
\showURL{%
\tempurl}


\bibitem[Ast(2017)]%
        {Asterius}
 \bibinfo{year}{2017}\natexlab{}.
\newblock \bibinfo{booktitle}{\emph{{Asterius: a Haskell to WebAssembly compiler}}}.
\newblock
\urldef\tempurl%
\url{https://github.com/tweag/asterius}
\showURL{%
\tempurl}


\bibitem[opt(2017)]%
        {optimization}
 \bibinfo{year}{2017}\natexlab{}.
\newblock \bibinfo{title}{{Code Optimizations in Emscripten}}.
\newblock
\newblock
\urldef\tempurl%
\url{https://emscripten.org/docs/optimizing/Optimizing-Code.html}
\showURL{%
\tempurl}


\bibitem[fil(2017)]%
        {filesystem}
 \bibinfo{year}{2017}\natexlab{}.
\newblock \bibinfo{title}{{File System Overview in Emscripten}}.
\newblock
\newblock
\urldef\tempurl%
\url{https://emscripten.org/docs/porting/files/file_systems_overview.html}
\showURL{%
\tempurl}


\bibitem[por(2017)]%
        {portability}
 \bibinfo{year}{2017}\natexlab{}.
\newblock \bibinfo{title}{{Portability At The C/C++ Level}}.
\newblock
\newblock
\urldef\tempurl%
\url{https://webassembly.org/docs/portability}
\showURL{%
\tempurl}


\bibitem[Was(2017)]%
        {Wasm-Bindgen}
 \bibinfo{year}{2017}\natexlab{}.
\newblock \bibinfo{booktitle}{\emph{{Wasm-Bindgen: a Rust library and CLI tool for facilitating high-level interactions between WebAssembly modules and JavaScript}}}.
\newblock
\urldef\tempurl%
\url{https://github.com/rustwasm/wasm-bindgen}
\showURL{%
\tempurl}


\bibitem[Che(2019)]%
        {Cheerp}
 \bibinfo{year}{2019}\natexlab{}.
\newblock \bibinfo{booktitle}{\emph{{Cheerp: an enterprise-grade C++ compiler for the web}}}.
\newblock
\urldef\tempurl%
\url{https://leaningtech.com/cheerp}
\showURL{%
\tempurl}


\bibitem[gra(2023)]%
        {graaf}
 \bibinfo{year}{2023}\natexlab{}.
\newblock \bibinfo{booktitle}{\emph{{Graaf lib: A general-purpose lightweight graph library}}}.
\newblock
\urldef\tempurl%
\url{https://bobluppes.github.io/graaf}
\showURL{%
\tempurl}


\bibitem[Battagline(2019)]%
        {battagline2019hands}
\bibfield{author}{\bibinfo{person}{Rick Battagline}.} \bibinfo{year}{2019}\natexlab{}.
\newblock \bibinfo{booktitle}{\emph{{Hands-On Game Development with WebAssembly: Learn WebAssembly C++ Programming by Building A Retro Space Game}}}.
\newblock \bibinfo{publisher}{Packt Publishing Ltd}.
\newblock


\bibitem[Boland and Black(2012)]%
        {6329885}
\bibfield{author}{\bibinfo{person}{Tim Boland} {and} \bibinfo{person}{Paul~E. Black}.} \bibinfo{year}{2012}\natexlab{}.
\newblock \showarticletitle{{Juliet 1.1 C/C++ and Java Test Suite}}.
\newblock \bibinfo{journal}{\emph{Computer}} \bibinfo{volume}{45}, \bibinfo{number}{10} (\bibinfo{year}{2012}), \bibinfo{pages}{88--90}.
\newblock
\urldef\tempurl%
\url{https://doi.org/10.1109/MC.2012.345}
\showDOI{\tempurl}


\bibitem[Cao et~al\mbox{.}(2023)]%
        {WRTester}
\bibfield{author}{\bibinfo{person}{Shangtong Cao}, \bibinfo{person}{Ningyu He}, \bibinfo{person}{Xinyu She}, \bibinfo{person}{Yixuan Zhang}, \bibinfo{person}{Mu Zhang}, {and} \bibinfo{person}{Haoyu Wang}.} \bibinfo{year}{2023}\natexlab{}.
\newblock \bibinfo{title}{{WRTester: Differential Testing of WebAssembly Runtimes via Semantic-aware Binary Generation}}.
\newblock
\newblock
\showeprint[arxiv]{2312.10456}
\urldef\tempurl%
\url{https://doi.org/10.48550/arXiv.2312.10456}
\showURL{%
\tempurl}


\bibitem[Chen et~al\mbox{.}(2022)]%
        {10.1145/3533767.3534218}
\bibfield{author}{\bibinfo{person}{Weimin Chen}, \bibinfo{person}{Zihan Sun}, \bibinfo{person}{Haoyu Wang}, \bibinfo{person}{Xiapu Luo}, \bibinfo{person}{Haipeng Cai}, {and} \bibinfo{person}{Lei Wu}.} \bibinfo{year}{2022}\natexlab{}.
\newblock \showarticletitle{{WASAI: Uncovering Vulnerabilities in Wasm Smart Contracts}}. In \bibinfo{booktitle}{\emph{Proceedings of the 31st ACM SIGSOFT International Symposium on Software Testing and Analysis}} \emph{(\bibinfo{series}{ISSTA '22})}. \bibinfo{pages}{703–715}.
\newblock
\showISBNx{9781450393799}
\urldef\tempurl%
\url{https://doi.org/10.1145/3533767.3534218}
\showDOI{\tempurl}


\bibitem[Cowan et~al\mbox{.}(1998)]%
        {cowan1998stackguard}
\bibfield{author}{\bibinfo{person}{Crispan Cowan}, \bibinfo{person}{Calton Pu}, \bibinfo{person}{Dave Maier}, \bibinfo{person}{Jonathan Walpole}, \bibinfo{person}{Peat Bakke}, \bibinfo{person}{Steve Beattie}, \bibinfo{person}{Aaron Grier}, \bibinfo{person}{Perry Wagle}, \bibinfo{person}{Qian Zhang}, {and} \bibinfo{person}{Heather Hinton}.} \bibinfo{year}{1998}\natexlab{}.
\newblock \showarticletitle{{{StackGuard}: Automatic Adaptive Detection and Prevention of {Buffer-Overflow} Attacks}}. In \bibinfo{booktitle}{\emph{7th USENIX Security Symposium (USENIX Security 98)}}, Vol.~\bibinfo{volume}{98}. \bibinfo{pages}{63--78}.
\newblock
\urldef\tempurl%
\url{https://www.usenix.org/conference/7th-usenix-security-symposium/stackguard-automatic-adaptive-detection-and-prevention}
\showURL{%
\tempurl}


\bibitem[Haas et~al\mbox{.}(2017)]%
        {WebAssemblyPLDI}
\bibfield{author}{\bibinfo{person}{Andreas Haas}, \bibinfo{person}{Andreas Rossberg}, \bibinfo{person}{Derek~L. Schuff}, \bibinfo{person}{Ben~L. Titzer}, \bibinfo{person}{Michael Holman}, \bibinfo{person}{Dan Gohman}, \bibinfo{person}{Luke Wagner}, \bibinfo{person}{Alon Zakai}, {and} \bibinfo{person}{JF Bastien}.} \bibinfo{year}{2017}\natexlab{}.
\newblock \showarticletitle{{Bringing the Web Up to Speed with WebAssembly}}. In \bibinfo{booktitle}{\emph{Proceedings of the 38th ACM SIGPLAN Conference on Programming Language Design and Implementation}} \emph{(\bibinfo{series}{PLDI '17})}. \bibinfo{pages}{185–200}.
\newblock
\showISBNx{9781450349888}
\urldef\tempurl%
\url{https://doi.org/10.1145/3062341.3062363}
\showDOI{\tempurl}


\bibitem[Hamidy et~al\mbox{.}(2020)]%
        {hamidy2020differential}
\bibfield{author}{\bibinfo{person}{Gilang Hamidy} {et~al\mbox{.}}} \bibinfo{year}{2020}\natexlab{}.
\newblock \showarticletitle{{Differential Fuzzing the WebAssembly}}.
\newblock  (\bibinfo{year}{2020}).
\newblock


\bibitem[He et~al\mbox{.}(2021)]%
        {272292}
\bibfield{author}{\bibinfo{person}{Ningyu He}, \bibinfo{person}{Ruiyi Zhang}, \bibinfo{person}{Haoyu Wang}, \bibinfo{person}{Lei Wu}, \bibinfo{person}{Xiapu Luo}, \bibinfo{person}{Yao Guo}, \bibinfo{person}{Ting Yu}, {and} \bibinfo{person}{Xuxian Jiang}.} \bibinfo{year}{2021}\natexlab{}.
\newblock \showarticletitle{{{EOSAFE}: Security Analysis of {EOSIO} Smart Contracts}}. In \bibinfo{booktitle}{\emph{30th USENIX Security Symposium (USENIX Security 21)}}. \bibinfo{pages}{1271--1288}.
\newblock
\showISBNx{978-1-939133-24-3}
\urldef\tempurl%
\url{https://www.usenix.org/conference/usenixsecurity21/presentation/he-ningyu}
\showURL{%
\tempurl}


\bibitem[Hilbig et~al\mbox{.}(2021)]%
        {WasmBinRealWorld}
\bibfield{author}{\bibinfo{person}{Aaron Hilbig}, \bibinfo{person}{Daniel Lehmann}, {and} \bibinfo{person}{Michael Pradel}.} \bibinfo{year}{2021}\natexlab{}.
\newblock \showarticletitle{{An Empirical Study of Real-World WebAssembly Binaries: Security, Languages, Use Cases}}. In \bibinfo{booktitle}{\emph{Proceedings of the Web Conference 2021}} \emph{(\bibinfo{series}{WWW '21})}. \bibinfo{pages}{2696–2708}.
\newblock
\showISBNx{9781450383127}
\urldef\tempurl%
\url{https://doi.org/10.1145/3442381.3450138}
\showDOI{\tempurl}


\bibitem[Hoque and Harras(2022)]%
        {10034550}
\bibfield{author}{\bibinfo{person}{Mohammed~Nurul Hoque} {and} \bibinfo{person}{Khaled~A. Harras}.} \bibinfo{year}{2022}\natexlab{}.
\newblock \showarticletitle{{WebAssembly for Edge Computing: Potential and Challenges}}.
\newblock \bibinfo{journal}{\emph{IEEE Communications Standards Magazine}} \bibinfo{volume}{6}, \bibinfo{number}{4} (\bibinfo{year}{2022}), \bibinfo{pages}{68--73}.
\newblock
\urldef\tempurl%
\url{https://doi.org/10.1109/MCOMSTD.0001.2000068}
\showDOI{\tempurl}


\bibitem[Jiang and Su(2009)]%
        {10.1145/1572272.1572283}
\bibfield{author}{\bibinfo{person}{Lingxiao Jiang} {and} \bibinfo{person}{Zhendong Su}.} \bibinfo{year}{2009}\natexlab{}.
\newblock \showarticletitle{Automatic mining of functionally equivalent code fragments via random testing}. In \bibinfo{booktitle}{\emph{Proceedings of the Eighteenth International Symposium on Software Testing and Analysis}} \emph{(\bibinfo{series}{ISSTA '09})}. \bibinfo{pages}{81–92}.
\newblock
\showISBNx{9781605583389}
\urldef\tempurl%
\url{https://doi.org/10.1145/1572272.1572283}
\showDOI{\tempurl}


\bibitem[Jiang et~al\mbox{.}(2023)]%
        {Warpdiff}
\bibfield{author}{\bibinfo{person}{Shuyao Jiang}, \bibinfo{person}{Ruiying Zeng}, \bibinfo{person}{Zihao Rao}, \bibinfo{person}{Jiazhen Gu}, \bibinfo{person}{Yangfan Zhou}, {and} \bibinfo{person}{Michael~R. Lyu}.} \bibinfo{year}{2023}\natexlab{}.
\newblock \showarticletitle{{Revealing Performance Issues in Server-Side WebAssembly Runtimes Via Differential Testing}}. In \bibinfo{booktitle}{\emph{Proceedings of the 38th IEEE/ACM International Conference on Automated Software Engineering}} \emph{(\bibinfo{series}{ASE '23})}. \bibinfo{pages}{661--672}.
\newblock
\urldef\tempurl%
\url{https://doi.org/10.1109/ASE56229.2023.00088}
\showDOI{\tempurl}


\bibitem[Kaluva and Hossain(2020)]%
        {kaluva2020webassembly}
\bibfield{author}{\bibinfo{person}{Lakshmi Venkata~Sainath Kaluva} {and} \bibinfo{person}{Abdullah Hossain}.} \bibinfo{year}{2020}\natexlab{}.
\newblock \bibinfo{title}{{WebAssembly for Video Analysis: An Explorative Multi-method study}}.
\newblock
\newblock


\bibitem[Khomtchouk(2021)]%
        {khomtchouk2021webassembly}
\bibfield{author}{\bibinfo{person}{Bohdan~B Khomtchouk}.} \bibinfo{year}{2021}\natexlab{}.
\newblock \showarticletitle{{WebAssembly Enables Low Latency Interoperable Augmented and Virtual Reality Software}}.
\newblock \bibinfo{journal}{\emph{arXiv preprint arXiv:2110.07128}} (\bibinfo{year}{2021}).
\newblock
\urldef\tempurl%
\url{https://doi.org/10.48550/arXiv.2110.07128}
\showURL{%
\tempurl}


\bibitem[Lehmann et~al\mbox{.}(2020)]%
        {255318}
\bibfield{author}{\bibinfo{person}{Daniel Lehmann}, \bibinfo{person}{Johannes Kinder}, {and} \bibinfo{person}{Michael Pradel}.} \bibinfo{year}{2020}\natexlab{}.
\newblock \showarticletitle{{Everything Old is New Again: Binary Security of {WebAssembly}}}. In \bibinfo{booktitle}{\emph{29th USENIX Security Symposium (USENIX Security 20)}}. \bibinfo{pages}{217--234}.
\newblock
\showISBNx{978-1-939133-17-5}
\urldef\tempurl%
\url{https://www.usenix.org/conference/usenixsecurity20/presentation/lehmann}
\showURL{%
\tempurl}


\bibitem[Li et~al\mbox{.}(2022)]%
        {10.1145/3498361.3538922}
\bibfield{author}{\bibinfo{person}{Borui Li}, \bibinfo{person}{Hongchang Fan}, \bibinfo{person}{Yi Gao}, {and} \bibinfo{person}{Wei Dong}.} \bibinfo{year}{2022}\natexlab{}.
\newblock \showarticletitle{{Bringing WebAssembly to Resource-constrained IoT Devices for Seamless Device-cloud Integration}}. In \bibinfo{booktitle}{\emph{Proceedings of the 20th Annual International Conference on Mobile Systems, Applications and Services}} \emph{(\bibinfo{series}{MobiSys '22})}. \bibinfo{pages}{261–272}.
\newblock
\showISBNx{9781450391856}
\urldef\tempurl%
\url{https://doi.org/10.1145/3498361.3538922}
\showDOI{\tempurl}


\bibitem[Li et~al\mbox{.}(2015)]%
        {10.1145/2786805.2786879}
\bibfield{author}{\bibinfo{person}{Ding Li}, \bibinfo{person}{Yingjun Lyu}, \bibinfo{person}{Mian Wan}, {and} \bibinfo{person}{William G.~J. Halfond}.} \bibinfo{year}{2015}\natexlab{}.
\newblock \showarticletitle{{String Analysis for Java and Android Applications}}. In \bibinfo{booktitle}{\emph{Proceedings of the 2015 10th Joint Meeting on Foundations of Software Engineering}} \emph{(\bibinfo{series}{ESEC/FSE '15})}. \bibinfo{pages}{661–672}.
\newblock
\showISBNx{9781450336758}
\urldef\tempurl%
\url{https://doi.org/10.1145/2786805.2786879}
\showDOI{\tempurl}


\bibitem[Li and Su(2023)]%
        {CompDiff}
\bibfield{author}{\bibinfo{person}{Shaohua Li} {and} \bibinfo{person}{Zhendong Su}.} \bibinfo{year}{2023}\natexlab{}.
\newblock \showarticletitle{{Finding Unstable Code via Compiler-Driven Differential Testing}}. In \bibinfo{booktitle}{\emph{Proceedings of the 28th ACM International Conference on Architectural Support for Programming Languages and Operating Systems, Volume 3}} \emph{(\bibinfo{series}{ASPLOS '23})}. \bibinfo{pages}{238–251}.
\newblock
\showISBNx{9781450399180}
\urldef\tempurl%
\url{https://doi.org/10.1145/3582016.3582053}
\showDOI{\tempurl}


\bibitem[Liu et~al\mbox{.}(2023)]%
        {10.1145/3618257.3624833}
\bibfield{author}{\bibinfo{person}{Kaiyan Liu}, \bibinfo{person}{Nan Wu}, {and} \bibinfo{person}{Bo Han}.} \bibinfo{year}{2023}\natexlab{}.
\newblock \showarticletitle{{Demystifying Web-based Mobile Extended Reality Accelerated by WebAssembly}}. In \bibinfo{booktitle}{\emph{Proceedings of the 2023 ACM on Internet Measurement Conference}} \emph{(\bibinfo{series}{IMC '23})}. \bibinfo{pages}{145–153}.
\newblock
\showISBNx{9798400703829}
\urldef\tempurl%
\url{https://doi.org/10.1145/3618257.3624833}
\showDOI{\tempurl}


\bibitem[McKeeman(1998)]%
        {McKeeman1998DifferentialTF}
\bibfield{author}{\bibinfo{person}{William~M. McKeeman}.} \bibinfo{year}{1998}\natexlab{}.
\newblock \showarticletitle{{Differential Testing for Software}}.
\newblock \bibinfo{journal}{\emph{Digit. Tech. J.}}  \bibinfo{volume}{10} (\bibinfo{year}{1998}), \bibinfo{pages}{100--107}.
\newblock
\urldef\tempurl%
\url{https://api.semanticscholar.org/CorpusID:14018070}
\showURL{%
\tempurl}


\bibitem[Mendki(2020)]%
        {9283720}
\bibfield{author}{\bibinfo{person}{Pankaj Mendki}.} \bibinfo{year}{2020}\natexlab{}.
\newblock \showarticletitle{{Evaluating Webassembly Enabled Serverless Approach for Edge Computing}}. In \bibinfo{booktitle}{\emph{2020 IEEE Cloud Summit}}. \bibinfo{pages}{161--166}.
\newblock
\urldef\tempurl%
\url{https://doi.org/10.1109/IEEECloudSummit48914.2020.00031}
\showDOI{\tempurl}


\bibitem[Moor et~al\mbox{.}(2007)]%
        {4362893}
\bibfield{author}{\bibinfo{person}{Oege~de Moor}, \bibinfo{person}{Mathieu Verbaere}, \bibinfo{person}{Elnar Hajiyev}, \bibinfo{person}{Pavel Avgustinov}, \bibinfo{person}{Torbjorn Ekman}, \bibinfo{person}{Neil Ongkingco}, \bibinfo{person}{Damien Sereni}, {and} \bibinfo{person}{Julian Tibble}.} \bibinfo{year}{2007}\natexlab{}.
\newblock \showarticletitle{{Keynote Address: .QL for Source Code Analysis}}. In \bibinfo{booktitle}{\emph{Proceedings of the 7th IEEE International Working Conference on Source Code Analysis and Manipulation}} \emph{(\bibinfo{series}{SCAM '07})}. \bibinfo{pages}{3–16}.
\newblock
\showISBNx{0769528805}
\urldef\tempurl%
\url{https://doi.org/10.1109/SCAM.2007.13}
\showDOI{\tempurl}


\bibitem[Ofenbeck et~al\mbox{.}(2016)]%
        {RandIR}
\bibfield{author}{\bibinfo{person}{Georg Ofenbeck}, \bibinfo{person}{Tiark Rompf}, {and} \bibinfo{person}{Markus P\"{u}schel}.} \bibinfo{year}{2016}\natexlab{}.
\newblock \showarticletitle{{RandIR: Differential Testing for Embedded Compilers}}. In \bibinfo{booktitle}{\emph{Proceedings of the 2016 7th ACM SIGPLAN Symposium on Scala}} \emph{(\bibinfo{series}{SCALA '16})}. \bibinfo{pages}{21–30}.
\newblock
\showISBNx{9781450346481}
\urldef\tempurl%
\url{https://doi.org/10.1145/2998392.2998397}
\showDOI{\tempurl}


\bibitem[Per\'{e}nyi and Midtgaard(2020)]%
        {perenyi2020stack}
\bibfield{author}{\bibinfo{person}{\'{A}rp\'{a}d Per\'{e}nyi} {and} \bibinfo{person}{Jan Midtgaard}.} \bibinfo{year}{2020}\natexlab{}.
\newblock \showarticletitle{{Stack-Driven Program Generation of WebAssembly}}. In \bibinfo{booktitle}{\emph{Programming Languages and Systems: 18th Asian Symposium, APLAS 2020}}. \bibinfo{publisher}{Springer-Verlag}, \bibinfo{pages}{209–230}.
\newblock
\showISBNx{978-3-030-64436-9}
\urldef\tempurl%
\url{https://doi.org/10.1007/978-3-030-64437-6_11}
\showDOI{\tempurl}


\bibitem[Pisanò and Servetti(2023)]%
        {10335388}
\bibfield{author}{\bibinfo{person}{Davide Pisanò} {and} \bibinfo{person}{Antonio Servetti}.} \bibinfo{year}{2023}\natexlab{}.
\newblock \showarticletitle{{Audio-aware Applications at the Edge Using in-browser WebAssembly and Fingerprinting}}. In \bibinfo{booktitle}{\emph{Proceedings of the 4th International Symposium on the Internet of Sounds}}. \bibinfo{pages}{1--9}.
\newblock
\urldef\tempurl%
\url{https://doi.org/10.1109/IEEECONF59510.2023.10335388}
\showDOI{\tempurl}


\bibitem[Ray(2023)]%
        {fi15080275}
\bibfield{author}{\bibinfo{person}{Partha~Pratim Ray}.} \bibinfo{year}{2023}\natexlab{}.
\newblock \showarticletitle{{An Overview of WebAssembly for IoT: Background, Tools, State-of-the-Art, Challenges, and Future Directions}}.
\newblock \bibinfo{journal}{\emph{Future Internet}} \bibinfo{volume}{15}, \bibinfo{number}{8} (\bibinfo{year}{2023}).
\newblock
\showISSN{1999-5903}
\urldef\tempurl%
\url{https://doi.org/10.3390/fi15080275}
\showDOI{\tempurl}


\bibitem[Romano et~al\mbox{.}(2022)]%
        {9678776}
\bibfield{author}{\bibinfo{person}{Alan Romano}, \bibinfo{person}{Xinyue Liu}, \bibinfo{person}{Yonghwi Kwon}, {and} \bibinfo{person}{Weihang Wang}.} \bibinfo{year}{2022}\natexlab{}.
\newblock \showarticletitle{{An Empirical Study of Bugs in WebAssembly Compilers}}. In \bibinfo{booktitle}{\emph{Proceedings of the 36th IEEE/ACM International Conference on Automated Software Engineering}} \emph{(\bibinfo{series}{ASE '21})}. \bibinfo{pages}{42–54}.
\newblock
\showISBNx{9781665403375}
\urldef\tempurl%
\url{https://doi.org/10.1109/ASE51524.2021.9678776}
\showDOI{\tempurl}


\bibitem[Sharma et~al\mbox{.}(2023)]%
        {RustSmith}
\bibfield{author}{\bibinfo{person}{Mayank Sharma}, \bibinfo{person}{Pingshi Yu}, {and} \bibinfo{person}{Alastair~F. Donaldson}.} \bibinfo{year}{2023}\natexlab{}.
\newblock \showarticletitle{{RustSmith: Random Differential Compiler Testing for Rust}}. In \bibinfo{booktitle}{\emph{Proceedings of the 32nd ACM SIGSOFT International Symposium on Software Testing and Analysis}} \emph{(\bibinfo{series}{ISSTA '23})}. \bibinfo{pages}{1483–1486}.
\newblock
\showISBNx{9798400702211}
\urldef\tempurl%
\url{https://doi.org/10.1145/3597926.3604919}
\showDOI{\tempurl}


\bibitem[Sti\'{e}venart et~al\mbox{.}(2022)]%
        {stievenart2022security}
\bibfield{author}{\bibinfo{person}{Quentin Sti\'{e}venart}, \bibinfo{person}{Coen De~Roover}, {and} \bibinfo{person}{Mohammad Ghafari}.} \bibinfo{year}{2022}\natexlab{}.
\newblock \showarticletitle{{Security Risks of Porting C Programs to WebAssembly}}. In \bibinfo{booktitle}{\emph{Proceedings of the 37th ACM/SIGAPP Symposium on Applied Computing}} \emph{(\bibinfo{series}{SAC '22})}. \bibinfo{pages}{1713–1722}.
\newblock
\showISBNx{9781450387132}
\urldef\tempurl%
\url{https://doi.org/10.1145/3477314.3507308}
\showDOI{\tempurl}


\bibitem[Stiévenart et~al\mbox{.}(2021)]%
        {9724846}
\bibfield{author}{\bibinfo{person}{Quentin Stiévenart}, \bibinfo{person}{Coen De~Roover}, {and} \bibinfo{person}{Mohammad Ghafari}.} \bibinfo{year}{2021}\natexlab{}.
\newblock \showarticletitle{{The Security Risk of Lacking Compiler Protection in WebAssembly}}. In \bibinfo{booktitle}{\emph{2021 IEEE 21st International Conference on Software Quality, Reliability and Security (QRS)}}. \bibinfo{pages}{132--139}.
\newblock
\urldef\tempurl%
\url{https://doi.org/10.1109/QRS54544.2021.00024}
\showDOI{\tempurl}


\bibitem[Tang et~al\mbox{.}(2023)]%
        {10.1145/3551349.3560432}
\bibfield{author}{\bibinfo{person}{Wei Tang}, \bibinfo{person}{Zhengzi Xu}, \bibinfo{person}{Chengwei Liu}, \bibinfo{person}{Jiahui Wu}, \bibinfo{person}{Shouguo Yang}, \bibinfo{person}{Yi Li}, \bibinfo{person}{Ping Luo}, {and} \bibinfo{person}{Yang Liu}.} \bibinfo{year}{2023}\natexlab{}.
\newblock \showarticletitle{{Towards Understanding Third-party Library Dependency in C/C++ Ecosystem}}. In \bibinfo{booktitle}{\emph{Proceedings of the 37th IEEE/ACM International Conference on Automated Software Engineering}} \emph{(\bibinfo{series}{ASE '22})}. Article \bibinfo{articleno}{106}, \bibinfo{numpages}{12}~pages.
\newblock
\showISBNx{9781450394758}
\urldef\tempurl%
\url{https://doi.org/10.1145/3551349.3560432}
\showDOI{\tempurl}


\bibitem[Wen and Weber(2020)]%
        {9156135}
\bibfield{author}{\bibinfo{person}{Elliott Wen} {and} \bibinfo{person}{Gerald Weber}.} \bibinfo{year}{2020}\natexlab{}.
\newblock \showarticletitle{{Wasmachine: Bring IoT up to Speed with A WebAssembly OS}}. In \bibinfo{booktitle}{\emph{2020 IEEE International Conference on Pervasive Computing and Communications Workshops (PerCom Workshops)}}. \bibinfo{pages}{1--4}.
\newblock
\urldef\tempurl%
\url{https://doi.org/10.1109/PerComWorkshops48775.2020.9156135}
\showDOI{\tempurl}


\bibitem[Yang et~al\mbox{.}(2011)]%
        {yang2011finding}
\bibfield{author}{\bibinfo{person}{Xuejun Yang}, \bibinfo{person}{Yang Chen}, \bibinfo{person}{Eric Eide}, {and} \bibinfo{person}{John Regehr}.} \bibinfo{year}{2011}\natexlab{}.
\newblock \showarticletitle{{Finding and Understanding Bugs in C Compilers}}. In \bibinfo{booktitle}{\emph{Proceedings of the 32nd ACM SIGPLAN Conference on Programming Language Design and Implementation}} \emph{(\bibinfo{series}{PLDI '11})}. \bibinfo{pages}{283–294}.
\newblock
\showISBNx{9781450306638}
\urldef\tempurl%
\url{https://doi.org/10.1145/1993498.1993532}
\showDOI{\tempurl}


\bibitem[Zhong(2023)]%
        {LeRe}
\bibfield{author}{\bibinfo{person}{Hao Zhong}.} \bibinfo{year}{2023}\natexlab{}.
\newblock \showarticletitle{{Enriching Compiler Testing with Real Program from Bug Report}}. In \bibinfo{booktitle}{\emph{Proceedings of the 37th IEEE/ACM International Conference on Automated Software Engineering}} \emph{(\bibinfo{series}{ASE '22})}. Article \bibinfo{articleno}{40}, \bibinfo{numpages}{12}~pages.
\newblock
\showISBNx{9781450394758}
\urldef\tempurl%
\url{https://doi.org/10.1145/3551349.3556894}
\showDOI{\tempurl}


\bibitem[Zhou and Chen(2023)]%
        {WASMOD}
\bibfield{author}{\bibinfo{person}{Jianfei Zhou} {and} \bibinfo{person}{Ting Chen}.} \bibinfo{year}{2023}\natexlab{}.
\newblock \showarticletitle{WASMOD: Detecting vulnerabilities in Wasm smart contracts}.
\newblock \bibinfo{journal}{\emph{IET Blockchain}} \bibinfo{volume}{3}, \bibinfo{number}{4} (\bibinfo{year}{2023}), \bibinfo{pages}{172--181}.
\newblock
\urldef\tempurl%
\url{https://doi.org/10.1049/blc2.12029}
\showDOI{\tempurl}


\bibitem[Zhou et~al\mbox{.}(2023)]%
        {WADIFF}
\bibfield{author}{\bibinfo{person}{Shiyao Zhou}, \bibinfo{person}{Muhui Jiang}, \bibinfo{person}{Weimin Chen}, \bibinfo{person}{Hao Zhou}, \bibinfo{person}{Haoyu Wang}, {and} \bibinfo{person}{Xiapu Luo}.} \bibinfo{year}{2023}\natexlab{}.
\newblock \showarticletitle{{WADIFF: A Differential Testing Framework for WebAssembly Runtimes}}. In \bibinfo{booktitle}{\emph{Proceedings of the 38th IEEE/ACM International Conference on Automated Software Engineering}} \emph{(\bibinfo{series}{ASE '23})}. \bibinfo{pages}{939--950}.
\newblock
\urldef\tempurl%
\url{https://doi.org/10.1109/ASE56229.2023.00188}
\showDOI{\tempurl}


\end{thebibliography}

\end{document}